\documentclass[apj]{emulateapj}
\newcommand{\myref}{./astro_references}
\usepackage{amsmath}
\usepackage{bm}

\shorttitle{Measuring Consistent Masses for 25 Milky Way Globular Clusters}
\shortauthors{}

\begin{document}
\title{Measuring Consistent Masses for 25 Milky Way Globular Clusters}
\author{Brian Kimmig\altaffilmark{1}, Anil Seth\altaffilmark{1},
Inese I. Ivans\altaffilmark{1}, Jay Strader\altaffilmark{2}, Nelson
Caldwell\altaffilmark{3}, Tim Anderton\altaffilmark{1}, Dylan Gregersen\altaffilmark{1}} 

\altaffiltext{1}{Physics and Astronomy Department, University of Utah, SLC, UT
84112}
\altaffiltext{2}{Department of Physics and Astronomy, Michigan State
University, East Lansing, Michigan 48824}
\altaffiltext{3}{Harvard-Smithsonian Center for Astrophysics, 60 Garden Street,
Cambridge, MA 02138, USA}

\begin{abstract}
We present central velocity dispersions, masses, mass to light ratios ($M/L$s), and rotation strengths for 25 Galactic globular clusters. We derive radial velocities of 1951 stars in 12 globular clusters from single order spectra taken with Hectochelle on the MMT telescope. To this sample we add an analysis of available archival data of individual stars.  For the full set of data we fit King models to derive consistent dynamical parameters for the clusters. We find good agreement between single mass King models and the observed radial dispersion profiles.  The large, uniform sample of dynamical masses we derive enables us to examine trends of $M/L$ with cluster mass and metallicity.  The overall values of $M/L$ and the trends with mass and metallicity are consistent with existing measurements from a large sample of M~31 clusters. This includes a clear trend of increasing $M/L$ with cluster mass, and lower than expected $M/L$s for the metal-rich clusters.  We find no clear trend of increasing rotation with increasing cluster metallicity suggested in previous work.
\end{abstract}

\section{Introduction}
Despite detailed knowledge of the chemistry and light profiles of globular clusters (GCs), their formation and evolution remains poorly understood. The kinematics of clusters play a key role in our understanding of GCs, most fundamentally as indicators of their mass and mass-to-light ratios ($M/L$s).  The $M/L$s of GCs reflects their initial mass function and their subsequent dynamical evolution.

The $M/L$ of clusters is typically determined by comparing central dispersion values or resolved stellar kinematics at a range of radii to a dynamical model constrained by the light profile of the cluster. The first large study of GC $M/L$s was by \cite{mandushev1991} who determined masses and $M/L$s for 32 galactic GCs.  Shortly thereafter, \cite{pryor1993} published the best referenced study of central velocity dispersions, masses, and $M/L$s for 56 galactic GCs. A more recent compilation of dynamical $M/L$s relying on literature measurements of the kinematics was presented by \cite{mclaughlin2005} for 57 clusters.  These studies all share in common the use of heterogeneous data sets, with kinematics derived from many sources and with a combination central dispersion measurements and individual stellar velocity measurements.  

The advent of multi-object spectrographs has enabled higher quality kinematics in large numbers of globular clusters.  Using an impressive data set of individual stellar velocities from the AAOmega spectrograph, \cite{lane2010b} derived cluster $M/L$ ratios in 10 halo globular clusters.  However, they chose not to use the light profile of clusters to constrain the cluster $M/L$s and fit both the kinematics and light profiles to separate Plummer models. This makes their results difficult to compare to previous work, which assumed single and multi-mass King models in deriving cluster $M/L$s.  Another recent study of cluster kinematics is by \cite{bellazzini2012}; this study includes central dispersion measurements but does not derive $M/L$s.  

In recent years, the number of extragalactic GCs with dynamical $M/L$s has eclipsed Galactic measurements.  The largest current study of cluster $M/L$s are from high resolution integrated dispersion measurements for 163~GCs in M31 \citep{strader2009,strader2011}.  There is some additional work measuring $M/L$s in local group galaxies \citep{djorgovski1997,larsen2002}, while studies in more distant galaxies have focused on the most massive GCs and ultra-compact dwarf galaxies \citep[e.g.][]{martini2004,hasegan2005,hilker2007,mieske2008,taylor2010}.  These studies typically derive King surface brightness profiles from high resolution $HST$ data and use these to create single-mass King dynamical profiles in deriving cluster masses.

These studies have revealed a positive correlation between $M/L$ and mass.  This trend was first seen by \cite{mandushev1991} and was verified in the larger sample of M31 cluster by \cite{strader2011}.  The increase of $M/L$ with increasing cluster mass can be explained by the preferential loss of low mass stars due to dynamical evolution \citep{kruijssen2009a}.  The timescale for mass loss in a cluster, the dissolution timescale, is related to its relaxation time which increases with increasing mass.  Thus, lower mass clusters undergo more low mass star loss and have reduced $M/L$s, while higher mass clusters ($\gtrsim 10^6 M_\odot$) preserve their initial mass functions.  The dissolution time scale is also dependent on the galactic environment in which the cluster lives \citep{kruijssen2008,kruijssen2009a}.

\cite{strader2011} find a trend of decreasing $M/L$ with increasing metallicity in their sample of 163 GCs in M31.  The trend is also visible in the M31~GC data of \cite{djorgovski1997}.  This trend is the opposite of the trend expected from simple stellar population models with a constant initial mass function \citep[IMF; e.g.][]{bruzual2003,maraston2005,conroy2010}.  Furthermore, it cannot be explained by dynamical effects: the metal-rich and metal-poor clusters sampled have similar sizes and masses, and the trend remains even at fixed galactocentric radius \citep{strader2011}.  One possibility is that this trend reflects a systematic change in the IMF, with the metal-rich clusters having a fewer low-mass stars than the metal-poor clusters.  This trend is apparently opposite to the change in IMF required to explain dwarf-sensitive line-strengths and dynamical masses in metal-rich elliptical galaxies \citep[e.g.][]{vandokkum2010,cappellari2012}.

Rotation in clusters has not been studied as systematically as dispersion profiles. Both \cite{lane2010b} and \cite{bellazzini2012} derive rotations from their samples of individual stars and find many clusters have dynamically significant rotation.  \cite{bellazzini2012} examine correlations of rotation with other GC parameters, and find a correlation of rotation with metallicity and horizontal branch morphology; there is no clear physical interpretation of these trends.  One complication with measuring rotation is that the strength of rotation depends significantly on the radii where the kinematics are measured.  This is seen clearly in dynamical models of rotating clusters \citep{fiestas2006,bianchini2013}.  Most recently, rotation in GCs was studied by examining the gradient of rotation in the center of clusters \citep{fabricius2014}; they find that the central rotation correlates with the overall ellipticity of the cluster as well as the dispersion and central density of the cluster.  

Stellar velocities in GCs have been used to search for the presence of intermediate-mass black holes and dark matter in clusters.  The presence of black holes or dark matter is inferred from changes in the apparent $M/L$ of a GC either at the center (indicating a black hole) or in the outer regions (indicating dark matter).  Several detections of intermediate-mass black holes have been claimed \citep[e.g.][]{gebhardt2005,lutzgendorf2011,jalali2012}, but these remain controversial \citep[e.g.][]{vandermarel2010,strader2012,denbrok2014}. The search for dark matter has been inconclusive \citep[e.g.][]{lane2010b,ibata2013}.  

In this paper, we focus on a uniform analysis of $M/L$s and rotation in a sample of 25~GCs.  This work is the first large study of cluster $M/L$s that relies solely on modern measurements of individual stellar velocities.  Furthermore, we analyze this data using single mass king models \cite{king1966c}, the same technique used in assessing $M/L$s of extragalactic clusters \citep[e.g][]{larsen2002,mieske2008,strader2011}.  This enables direct comparison between our work and the largest available sample of high-quality $M/L$ measurements of GCs \citep{strader2011}.  

The paper is structured as follows. In Section \ref{sec:data} we discuss the data and its source. We then detail our techniques for obtaining average velocities, central velocity dispersions, mass and $M/L$, and rotations for each cluster in Section \ref{sec:analysis}. Section \ref{sec:results} discusses our results and our agreement with previous work. Finally in Section \ref{sec:discussion} we discuss our results in the context of other large studies.

\section{Data}
\label{sec:data}

\subsection{Hectochelle}
The spectroscopic data acquired for this project were gathered with Hectochelle at the MMT observatory on Mount Hopkins, Arizona. Hectochelle \citep{szentgyorgyi2011} is a multi-object spectrograph, with a 1 degree field of view, that uses 240 1.5\arcsec\ diameter fibers to take single order spectra at resolution $\sim$38,000. The full spectral coverage of the instrument is 3800 - 9000 \AA. Observing is done with an order selecting filter that enables multi-object observations of a $\sim150$ \AA wide spectral range \citep{szentgyorgyi2011}. 

Target stars were selected from several sources of photometry. The primary source was the Sloan Digital Sky Survey Data Release 8 (SDSS DR8; \cite{aihara2011}). The centers of Milky Way globular clusters are excluded from SDSS pipeline photometry, but many clusters had their SDSS imaging analyzed separately by \cite{an2008}. These catalogs were merged with the SDSS stars from the outer regions of the clusters to make master photometric catalogs for six objects: M2, M3, NGC 4147, NGC 5053, and Pal 5. The cluster NGC 6934 is in SDSS DR8 but was not in earlier releases, so had no photometry from \cite{an2008}; we manually photometered these data using DAOPHOT (Stetson 1987), using stars in the cluster outskirts with SDSS DR8 photometry for calibration. A final target, M14, was not in SDSS at all. Because it has significant foreground extinction, we used 2MASS to select red giants to target for spectroscopy.

For the clusters with SDSS photometry, we selected spectroscopic targets around a cluster fiducial sequence in a $g-i$ vs.$i$ color-magnitude diagram ($J-K$ vs.~$K$ was used for M14). While brighter stars were generally given higher priority in target selection, for each setup the target stars were restricted to a maximum range of $\sim 3.5$--4 mag because of scattered light. In practice, this often led to the exclusion of stars within about 1 mag of the tip of the red giant branch, depending on the distance to the cluster.

Because the Hectochelle fibers are placed into position on a magnetic plate, rather than a plug plate, the fibers lie on the focal plane itself, preventing very dense packing of fibers. For this reason, and to avoid problems with background subtraction, stars in the most central regions of the clusters are poorly represented in our samples. Table ~\ref{tab:ObsInfo} lists the numbers of stars observed for each cluster.

Our data were taken through the RV31 filter which has a central wavelength at 5230 \AA. The region covered by the RV31 filter includes the Mg b triplet (5167-5184 \AA) making it possible to determine velocities, even for the lower signal to noise [S/N] spectra. Reliable velocities were obtained for stars brighter than 21 magnitude in g. Therefore, the most distant target in our sample was NGC~6402 (M~14), which has a HB magnitude of 17.3. Considering all GCs closer than this most distant cluster and above declination of -10 degrees, our sample includes 12 out of 19 observable clusters.  This selection was due to the targets that were observable on the nights we took data. Our data had a typical S/N ratio per pixel of $\sim10$, and a mean velocity accuracy of $\sim0.5$ km/s. 

The spectra were reduced in a variation of the pipeline described in ~\cite{caldwell2009}. The final spectrum is sky subtracted. An error spectrum including the noise from both source and sky is also produced. We list in Table ~\ref{tab:ObsInfo} the observing information for each cluster, including the cluster name, observation date, number of stars observed on the run, exposure time, the total number of stars observed in the cluster, and the total number within the tidal radius. Data on Palomar 5 was also obtained, but is not analyzed here due to difficulty in membership selection and the corresponding lack of constraints obtained on the mass. We obtained spectra of 100 to 500 stars around each cluster. The median number of stars inside the tidal radius is 88, while the median number of members is 72 (see Section \ref{sec:analysis} for details).

\begin{deluxetable*}{lccccc}
\tabletypesize{\scriptsize}
\tablecolumns{6} 
\tablewidth{0pt}
\tablecaption{Observational Information \label{tab:ObsInfo}}
\tablehead{
   \colhead{Cluster} & \colhead{Obs. Date} & \colhead{Number Obs.} & \colhead{Exp. Time (sec)} & \colhead{$N_{Unique}$} & \colhead{$N_{in}$}}

\startdata
   M 2      & 2013 Sep 24  & 99  & 3@2400 &    & \\
            & 2013 Sep 30  & 99  & 3@2400 & 99 & 91 \\
   \\         
   M 3      & 2013 Apr 27  & 178 & 3@2700 & 178 & 157\\
   \\
   M 5      & 2013 Apr 28  & 204 & 3@2700 & 204 & 155\\
   \\
   M 14     & 2012 Jun 02   & 189 & 4@2700 &     & \\
            & 2012 Jun 07   & 192 & 2@2700 &     & \\
            & 2012 Jul 06   & 192 & 2@2700 & 340 & 103\\
   \\
   M 15     & 2011 Sep 14  & 108 & 4@2400 &     &\\
   		    & 2012 Nov 28 & 108 & 3@1925 & 108  & 84\\
   \\		
   M 53     & 2012 Apr 04   & 166 & 4@2700 & 166 & 110\\
   \\
   M 71     & 2013 Oct 01  & 139 & 4@1800 &     & \\
            & 2013 Dec 03  & 134 & 2@1800 & 140 & 91\\
   \\
   M 92     & 2012 May 09   & 92  & 3@2400 & 92 & 85 \\
   \\
   NGC 4147 & 2012 Feb 06   & 146 & 3@2700 &     & \\
   			& 2012 Apr 08   & 141 & 3@2700 &     & \\
   			& 2012 Apr 09   & 147 & 2@2700 & 148 & 15\\
   \\	
   NGC 5053 & 2012 Feb 04   & 133 & 5@2400 &     & \\
   			& 2012 Feb 06   & 133 & 5@2400 &     & \\
   			& 2013 Mar 26  & 116 & 5@2700 &      & \\
   			& 2013 Apr 23  & 116 & 5@2700 & 186  & 50\\
   \\			
   NGC 5466 & 2012 Feb 07   & 160 & 2@2400 &     &\\
            & 2012 Feb 09   & 160 & 2@2700 &     &\\
            & 2013 Feb 24  & 41  & 2@3600 &      &\\
            & 2013 Mar 26  & 28  & 3@2700 &      &\\
            & 2013 Apr 23  & 24  & 2@2700 &      &\\
            & 2013 Apr 24  & 70  & 2@2700 &      &\\
            & 2013 Apr 26  & 72  & 3@2700 &      &\\
            & 2013 Apr 28  & 46  & 2@2700 &      &\\
            & 2013 Jun 19  & 36  & 4@1200 &      &\\
            & 2013 Jun 20  & 20  & 4@1200 &      &\\
            & 2013 Jun 21  & 14  & 4@1200 &      &\\
            & 2013 Jun 22  & 20  & 4@1200 &      &\\
            & 2013 Jun 26  & 18  & 4@1200 &      &\\
            & 2013 Jun 28  & 13  & 2@1200 & 524  & 51\\
   \\

   NGC 6934 & 2012 Sep 01   & 185 & 3@2400 &     &\\
   			& 2012 Oct 13   & 185 & 3@1200 & 185   & 48\\
   \\			
\enddata
\end{deluxetable*}

\subsection{Archival}
Complimenting our Hectochelle data we have included an analysis of archival data. We examined previous literature where a large number of single star velocities are published. The archival data used in our study comes from numerous sources, including other multi-object spectrographs, each having velocity precision to $\sim1$ km/s \citep{carretta2007a, carretta2007c, carretta2007d, carretta2010, carretta2011, carretta2013a, carretta2013b, lane2011, drukier1998, drukier2007}.Table \ref{tab:archival} lists archival data used in this analysis, for more information on their reduction see the paper listed in column 3. The most common cluster name can be found in the first column, the second column contains the alternate identification of the cluster. Column 4 shows the code used to reference the dataset in the rest of the paper. Multiple data sets exist for seven of the 25 clusters.  We analyzed these data sets independently to check for consistency.

\begin{deluxetable}{llll}
\tablecolumns{4}
\tablewidth{0pt} 
\tabletypesize{\scriptsize}
\tablecaption{Archival Data \label{tab:archival}}
\tablehead{
   \colhead{Cluster} & \colhead{Alt. Id} & \colhead{Paper} & \colhead{Code} }
\startdata
   NGC 362  & \nodata & \cite{carretta2013b} & C13      \\
   NGC 2808 & \nodata & \cite{carretta2011}  & G11      \\
   M 5      & NGC 5904  & \cite{carretta2013a} & G13  \\
   M 12     & NGC 6218 & \cite{carretta2007a} & C07b \\     
   NGC 6441 & \nodata & \cite{carretta2007c} & G07      \\ 
   M 54     & NGC 6715 & \cite{carretta2010}  & C10  \\
   NGC 6752 & \nodata & \cite{carretta2007d} & C07     \\
   47 Tuc   & NGC 104 & \cite{carretta2013a} & G13  \\
   47 Tuc   & NGC 104 & \cite{lane2011} & L11 \\
   NGC 288  & \nodata  &  \cite{lane2011} & L11 \\ 
   M 68     & NGC 4590 &  \cite{lane2011} & L11  \\
   M 53     & NGC 5024 &  \cite{lane2011} & L11 \\      
   M 4      & NGC 6121  &   \cite{lane2011} & L11 \\
   M 12     & NGC 6218 &  \cite{lane2011} & L11 \\
   M 22     & NGC 6656 &  \cite{lane2011} & L11 \\
   NGC 6752 & \nodata  &  \cite{lane2011} & L11 \\
   M 55     & NGC 6809 &  \cite{lane2011} & L11 \\
   M 30     & NGC 7099 &  \cite{lane2011} & L11 \\
   M 92     & NGC 6341 & \cite{drukier2007} & D07 \\
   M 15     & NGC 7078 & \cite{drukier1998} & D98 \\
\enddata

\end{deluxetable}

\section{Analysis}
\label{sec:analysis}

\subsection{Radial Velocities and Errors}

Heliocentric radial velocities were derived using the XCSAO routine in IRAF's \citep{tody1993} RVSAO package \citep{kurtz1998} for all Hectochelle data. To ensure the best possible velocity determinations even at low S/N, we used the 30 \AA \ section, 5160-5190 \AA, for cross-correlation. This section included the strong Mg~b triplet lines, allowing better radial velocities to be obtained from lower signal to noise data. 

We use a Monte Carlo method to obtain a more robust estimate of the errors on each velocity than given by XCSAO.  The XCSAO errors are determined from a combination of the full width at half maximum of the correlation peak and the $R$ value to derive the error \citep{kurtz1998}. To derive velocity errors we first add Gaussian random noise to the spectrum based on the error spectrum, which is propagated through the reduction pipeline and includes the errors from Poisson, sky, and read noise. We then redetermine the velocity in the noise-added spectrum with XCSAO. This procedure is repeated 50 times for each stellar spectrum. We determine the velocity error from the standard deviation of all 50 measurements. XCSAO's velocity error estimates remain very low even at low S/N and don't capture the true uncertainty in low S/N measurements. We do however make use of the strength of cross-correlation; the $R$ value from XCSAO provides useful information on the confidence of the measurement that complements our Monte Carlo error determinations.

\begin{figure}
\centering
\epsscale{1.25}
\plotone{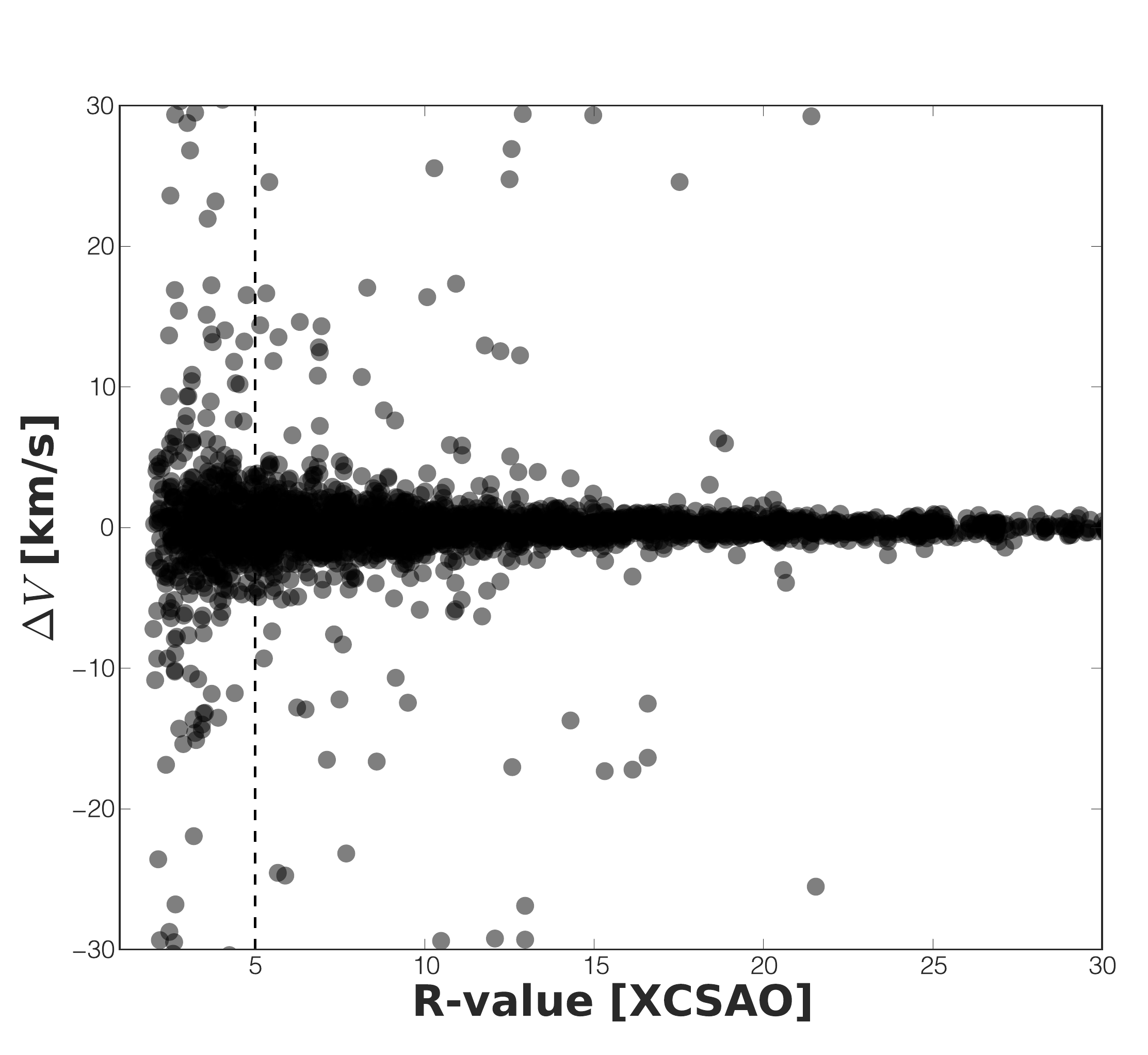}
\caption{Velocity difference between stars observed on more than one occasion by Hectochelle \citep{szentgyorgyi2011} versus the R value returned by XCSAO from cross correlation. Each difference is plotted twice; once at each measurement's $R$ value. The $R$ value is an indicator of the confidence in the measurement. The dashed vertical line at 5 indicates the cut we impose to compile our final velocity catalogues, below this we do not trust the measurements. For discrepant velocities (difference~$>$~2~km/s) with $R$ value $>5$, the velocity with the higher $R$ value is used in the final catalogue. In the case where velocities agree we combine measurements, averaging velocities and $R$ values and propogating errors for the final catalogue. } \label{fig:diffRVal}
\end{figure}

We verify that our error estimates are accurate by comparing repeat measurements of 1554 stars observed on different nights within our sample. Figure \ref{fig:diffRVal} shows the velocity difference of each repeat observation against the $R$ value of each night (thus each velocity difference is plotted twice). This figure shows that a large number of catastrophic outliers are present at $R$ values less than 5; thus we only analyze stars above this limit where we believe we can get reliable velocity measurements. Outliers at higher $R$ values may be due to binary stars, and based on a visual inspection of the spectra many of these objects appear to be foreground dwarfs which are expected to have a higher binary fraction than cluster stars.

\begin{figure}
\centering
\epsscale{1.25}
\plotone{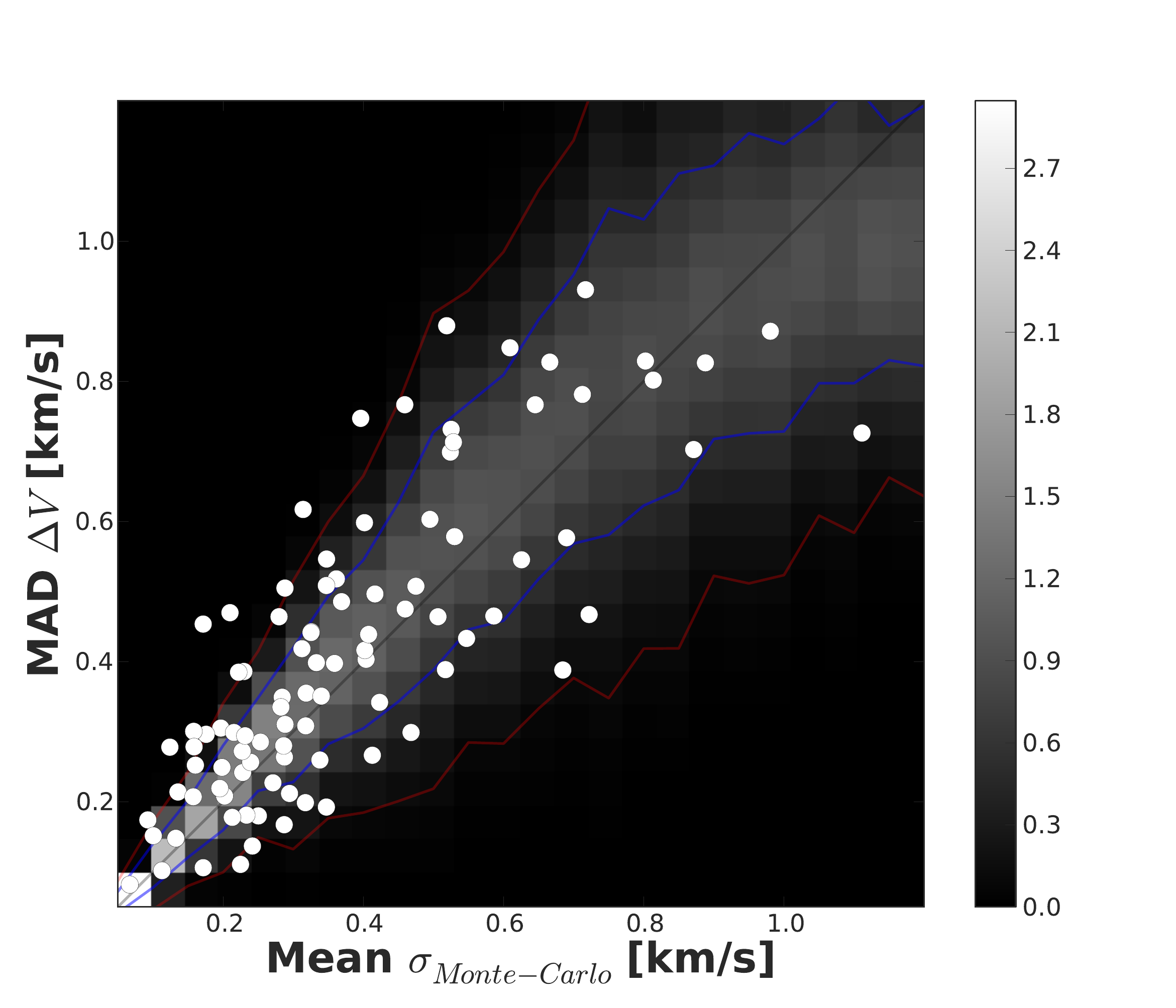}
\caption{The mean of the Monte Carlo errors plotted against the scaled median absolute deviations (MAD) of the night to night differences  (white points).  We scale the MAD by 1.4826 and divide by the square root of 2 to make it a more accurate estimate of the velocity errors. Bins  are defined by groupings of 20 $R$ values. The data is expected to follow the one-to-one line shown if our Monte Carlo errors are correctly estimated. In the background we show the simulations of the expected scatter in the measurement including an assumed binary fraction of 15\%. The 68\% and 95\% confidence intervals of the simulation are shown by the blue and red lines respectively. The overall scatter of the data is consistent with the expected scatter, while inclusion of a modest binary fraction is required to explain the bias towards higher MAD values.  This comparison gives us confidence that our velocity error estimates are robust. } 
\label{fig:madPlot}
\end{figure}

We can also use the repeat measurements to check the accuracy of our Monte Carlo errors. We check if our errors are over- or under-estimated by looking at the median absolute deviation (MAD) of the velocity differences against the mean of the Monte Carlo errors. With the correct scale factor the MAD is a reliable estimation of the standard deviation.  Assuming our errors are normally distributed the correct scale factor is 1.4826 \citep{birkes1993}.  We also divide the MAD values by the square root of two to account for the larger uncertainty in the subtracted difference compared to error estimates on a single measurement.  For correctly estimated errors, the scaled MAD should correlate roughly 1:1 with the mean of the Monte Carlo errors. In Figure \ref{fig:madPlot} we show this comparison for data binned by groupings of 20 $R$ values.  The scatter seen in the data about the 1:1 line is biased toward higher MAD values.  This suggests the binary fraction in our bins is not negligible, thus causing larger than expected deviations in the repeat measurements and biasing our MAD values. In the background we show the probability distribution for 100 MAD simulations in each $\sigma_{Monte-Carlo}$ bin.  For the simulations we assume the velocity measurements in each bin have Gaussian noise based on their average Monte Carlo error.  We can account for the bias of MAD values above the 1:1 line by including a 15\% binary fraction, where binary stars are given an additional velocity dispersion term of 50~km/s; the simulations are not sensitive to the exact value of this dispersion as long as it is significantly larger than the Monte Carlo error. The 68\% and 95\% confidence intervals are shown by the blue and red lines respectively. This comparison shows us that the scatter of the data points is comparable to the scatter expected if our errors are correctly estimated and that binary stars can account for the small bias we see towards higher MAD values.  This gives us confidence that our velocity error estimates are robust.

To create the final velocity catalogue of each cluster we impose an $R$ value cut, accepting velocities with $R$ greater than 5.  Our results are not sensitive to the precise value of this $R$ value cut, for instance using a minimum $R$ value of 10 changes our results by $< 1 \sigma$ for all cases where we have sufficient data.  For stars with repeat measurements, we average values if their velocities are within 2~km/s of each other, otherwise the velocity with the highest $R$ value is taken as the final velocity.  In combining velocities, errors are propagated through to determine the final error. We include all positional information along with the final combined individual stellar velocities and errors for stars used in this study in Table \ref{tab:hect_vel} of Appendix \ref{sec:vel_appendix}.

\subsection{Determining Cluster Velocity Dispersions}
In this study we focus on modeling the kinematic data of individual stars in a cluster with a single mass King model \cite{king1966c} (K66) to obtain central velocity dispersions. To do this, we use a variation to the iterative method introduced by \cite{pryor1993}. Specifically, we add a determination of cluster membership probabilities in our iterative method. With minor modifications, we use the technique described by \cite{walker2009}. 

To summarize our method, we start the iterative expectation maximization (EM) technique with an initial guess for the membership probability of each star and the variance in the velocity. During each iteration the average velocity, dispersion, and membership probabilities are recalculated. The central velocity dispersions are obtained by scaling a K66 model. We continued iterating until the parameters converged, i.e.~when the likelihood function was maximized.   We now explain our method in detail.

We use the EM technique to maximize the expected log likelihood (Equation \ref{eq:loglike}; see also Eq.~4 from \cite{walker2009}) and estimate the average velocity ($\langle \hat{V} \rangle$) and the central dispersion ($\hat{\sigma}_{0}$) using the radius ($a_i$), velocity ($V_{i}$) and velocity error ($\sigma_{V_{i}}$) of each star (later in the text $\langle \hat{V} \rangle$ is simplified to $\overline{V}$ and $\hat{\sigma}_{0}$ is simplified to $\sigma_{0}$).

\begin{equation}
\begin{split}
\label{eq:loglike}
E(\ln \ L (\langle \hat{V} \rangle , \hat{\sigma}_{0}^{2}) 
\ | \ S ) = 
\sum_{i=1}^{N} \hat{P}_{M_{i}} \ln [ \hat{p}_{mem}(V_{i}) \hat{p}(a_{i})] 
\\ 
+ \sum_{i=1}^{N} ( 1 - \hat{P}_{M_{i}} ) \ln [ \hat{p}_{non}(V_{i})[1 - 
\hat{p}(a_{i})]]
\end{split}
\end{equation}

$S$ is the full data set of $\{V_{i}, a_{i} \}$, $\hat{p}_{mem}(V_i)$ and $\hat{p}_{non}(V_i)$ are the membership and non-membership probabilities of each star, while $\hat{P}_{M_{i}}$ is the normalized membership probability which includes an {\em a priori} radial decrease in membership probability ($\hat{p}(a_i)$) as shown in Eq.~\ref{eq:membership}.  

Maximization of this likelihood function is obtained by iterating Eqs.~\ref{eq:normed_membership} to \ref{eq:variance} below. These are similar to the equations presented in \cite{walker2009}. However the dispersion expected at a given radius is given by $\hat{\sigma}_{0} \times \sigma_{K_{i}}$, where $\sigma_{K_{i}}$ represents a K66 model with the core radius and concentration taken from \cite{harris1996} (2010 edition) and a central dispersion of 1~km/s. The majority of the structural parameters used in this study, compiled in \cite{harris1996} (2010 edition), come from \cite{mclaughlin2005}.   

\begin{equation}\label{eq:normed_membership}
\begin{split}
\hat{P}_{M_{i}} = \\
&\frac{ 
\hat{p}_{mem}(V_{i}) 
\hat{p}(a_{i}) 
}
{
\hat{p}_{mem}(V_i) 
\hat{p}(a_{i}) 
+
\hat{p}_{non}(V_i) 
[1 - \hat{p}(a_{i})] 
}
\end{split}
\end{equation}

\begin{equation} \label{eq:membership}
\hat{p}_{mem}(V_i) = 
\frac{ \exp 
\left[ -\frac{1}{2} \left(
\frac{ [V_i - \langle \hat{V} \rangle]^2}
{(\hat{\sigma}_{0}\sigma_{K_{i}})^{2} + \sigma_{V_{i}}^2)} \right)
\right] }
{\sqrt{2\pi((\hat{\sigma}_{0}\sigma_{K_{i}})^{2} + \sigma_{V_{i}}^2)}}
\end{equation}

\begin{equation}\label{eq:nonmember}
\begin{split}
&\hat{p}_{non}(V_i) = \frac{1}{N_{bes}} \\ 
& \times \sum_{j=1}^{N} \frac{1}{\sqrt{2\pi\sigma_{V_{bes}}^{2}}} 
\exp \left[ -\frac{1}{2} \frac{(V_{bes_{j}} - V_i)^2}{\sigma_{V_{bes}}^{2}} \right]
\end{split}
\end{equation}

\begin{equation}\label{eq:p_a}
\hat{p}(a_{i}) = 
\min_{1 \leq u \leq i} 
\left[ 
\max_{i \leq v \leq N} 
\frac{\sum_{j=u}^{v} \hat{P}_{M_{j}}} 
{v - u +1} 
\right]
\end{equation}

\begin{equation}\label{eq:velocity}
\langle \hat{V} \rangle  = 
\frac{\sum_{i=1}^{N}
\frac{\hat{P}_{M_{i}} V_i }
{1 + \sigma_{V_{i}}^2 / ( \hat{\sigma}_{0} \sigma_{K_{i}} )^2} } 
{\sum_{i=1}^{N} 
\frac {\hat{P}_{M_{i}}}
{1 + \sigma_{V_{i}}^2 / (\hat{\sigma}_{0} \sigma_{K_{i}} )^2} }
\end{equation}

\begin{equation}\label{eq:variance}
\hat{\sigma}_{0}^2 = 
\frac{\sum_{i=1}^{N}
\frac{\hat{P}_{M_{i}} [V_i - \langle \hat{V} \rangle ]^2 }
{(1 + \sigma_{V_{i}}^2 / ( \hat{\sigma}_{0} \sigma_{K_{i}})^2 )^2 } } 
{\sum_{i=1}^{N} 
\frac {\hat{P}_{M_{i}} \sigma_{K_{i}}^2 }
{1 + \sigma_{V_{i}}^2 / (\hat{\sigma}_{0} \sigma_{K_{i}})^2 }}
\end{equation}

The final membership probability $\hat{P}_{M_{i}}$ (Eq.~\ref{eq:normed_membership}) is determined from three separate membership estimates. These estimates include information about the radius and velocity of the observed stars as well as the Besan\c{c}on model estimates of the expected foreground \& background \citep{robin2003}. The membership probability ($\hat{p}_{mem}(V_i)$) is determined assuming member stars follow a Gaussian velocity distribution at each radius centered on $\langle \hat{V} \rangle$ and with a width $\hat{\sigma}_{0}\sigma_{K_{i}}$. The probability of non-membership ($\hat{p}_{non}(V_i)$) incorporates information from the Besan\c{c}on model.  Each Besan\c{c}on model is generated with a 1 degree field of view centered on the cluster. The number of stars in this model is given by $N_{bes}$, while their velocities are $V_{bes_j}$. The Besan\c{c}on velocities are smoothed by a parameter $\sigma_{V_{bes}}$; we choose a value of 20 km/s to ensure proper estimation for the membership of stars in clusters with velocities close to the Galactic field. Finally, a radial decrease in membership probability is obtained by sorting our data radially from smallest to largest using and enforcing a steady decrease in probability based on the current estimates of $\hat{P}_{M_{i}}$ as shown in Equation~\ref{eq:p_a} \citep{grotzinger1984}. Once values of $\hat{P}_{M_{i}}$ are determined, equations \ref{eq:velocity} and \ref{eq:variance} are used to update the values in the parameter set $\{\langle \hat{V} \rangle , \hat{\sigma}_{0}^{2}\}$ at each iteration.

We initialize our membership probabilities $\hat{p}(a_{i}) = 0.5$ and $\hat{P}_{M_{i}} = 0.5$ for all $i$, and make an initial guess for $\hat{\sigma}_{0}^{2} = 4$ km/s.  We find, as \cite{walker2009} did, the the final parameters  are insensitive to the initialization of $\hat{\sigma}_{0}^{2}$.  After initializiation, we evaluate Equation \ref{eq:velocity}, giving us the initial parameters we needed to start iterations.  The full system of equations \ref{eq:normed_membership} to \ref{eq:variance}  are then iterated to determine a final $\langle \hat{V} \rangle$ and $\hat{\sigma}_{0}^{2}$.  Convergence  is obtained within 20 iterations; we  use 50 iterations for our final parameter estimates.

We calculate errors on the average velocity and central dispersion via bootstrapping. The dataset is randomly sampled and put through the complete  algorithm 100 times. We take the standard deviation of the 100 average velocities and dispersions as their respective errors.  We note that the ``sampling'' error determined from the bootstrapping is always larger than the random error determined from varying individual star velocities.  

\begin{figure}
\centering
\epsscale{1.25}
\plotone{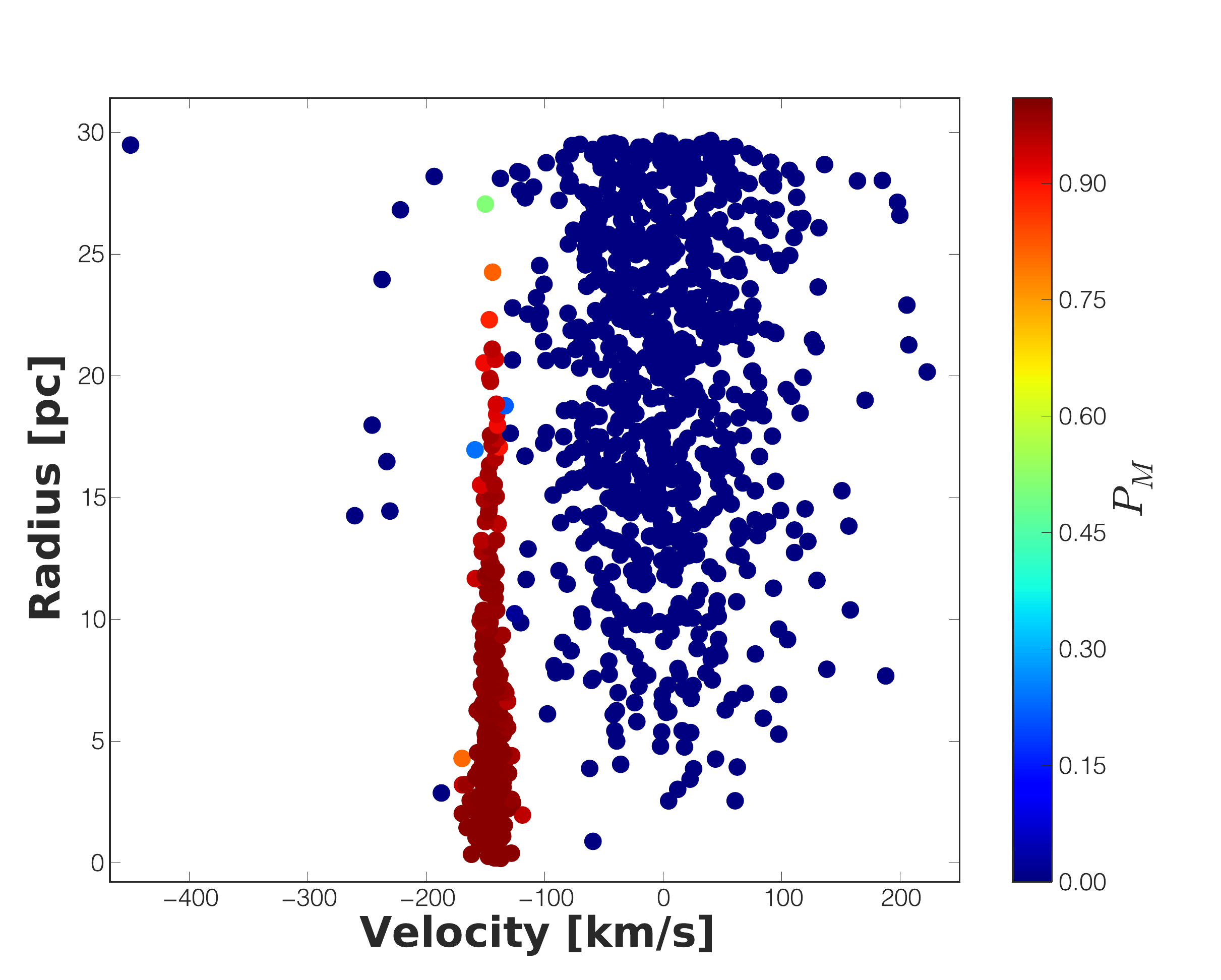}
\caption{Radius versus velocity for the cluster M 22 (NGC 6656). Coloring indicates the final cluster membership probability $P_M$ shown in Eq. \ref{eq:normed_membership}.  The membership probability is determined assuming the cluster stars have a Gaussian velocity distribution and that their membership probabilities monotonically decrease with radius. A model of the velocities of Galactic foreground and background stars is also incorporated. }\label{fig:m22Selection}
\end{figure}

Figure \ref{fig:m22Selection} illustrates the ability of the EM technique to select GC members from a crowded field; it shows the velocity plotted against the radius for stars in the dataset for the GC M~22. Each point is colored by its final membership probability. In addition to the membership peak around the cluster velocity, the radial decrease enforced by our membership probability assignment in Eq.~\ref{eq:p_a} is clearly seen.  

\begin{figure}
\centering
\epsscale{1.25}
\plotone{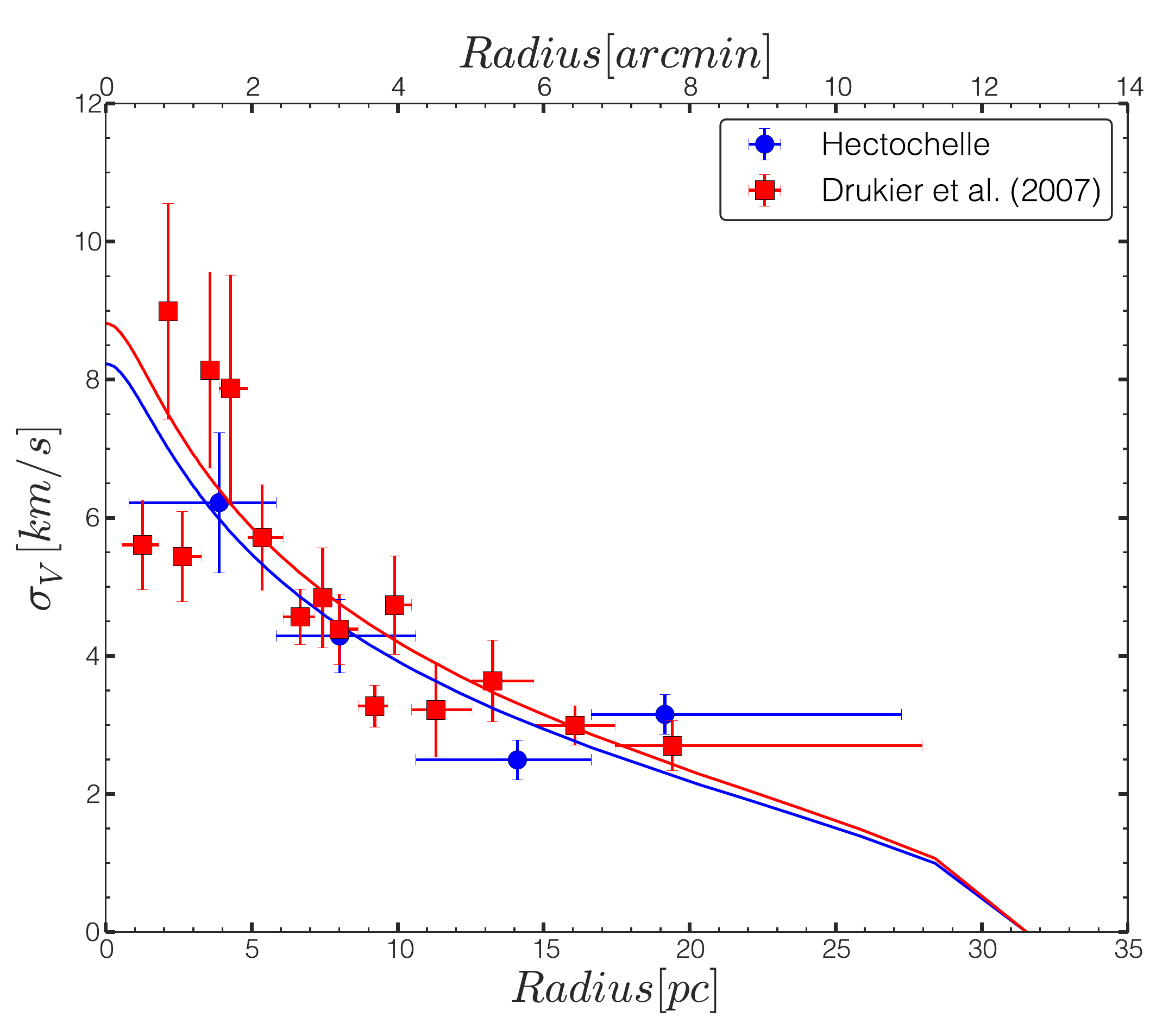}
\caption{The best fit dispersion profile is plotted above in solid lines for the GC M 92. Overplotted are dispersions derived from radially binned data; blue circles represent Hectochelle data and red squares represent \cite{drukier2007} data. Both data sets clearly fit their model well. We also see very good agreement between the two data sets. Bins are defined by radially grouping 20 stars.}\label{fig:m92radialdisp}
\end{figure}

\begin{figure*}
\centering
\epsscale{1.25}
\plotone{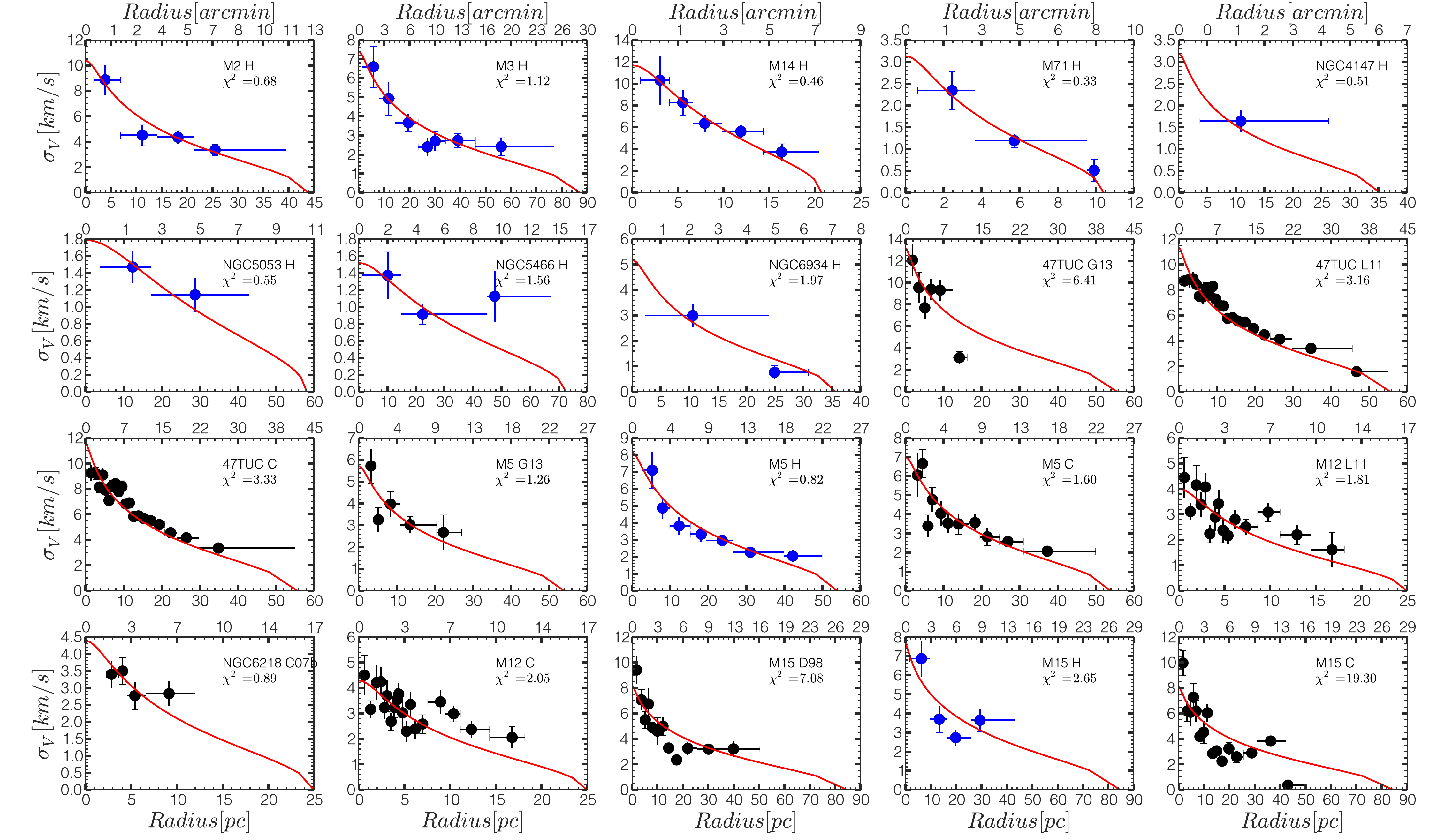}
\plotone{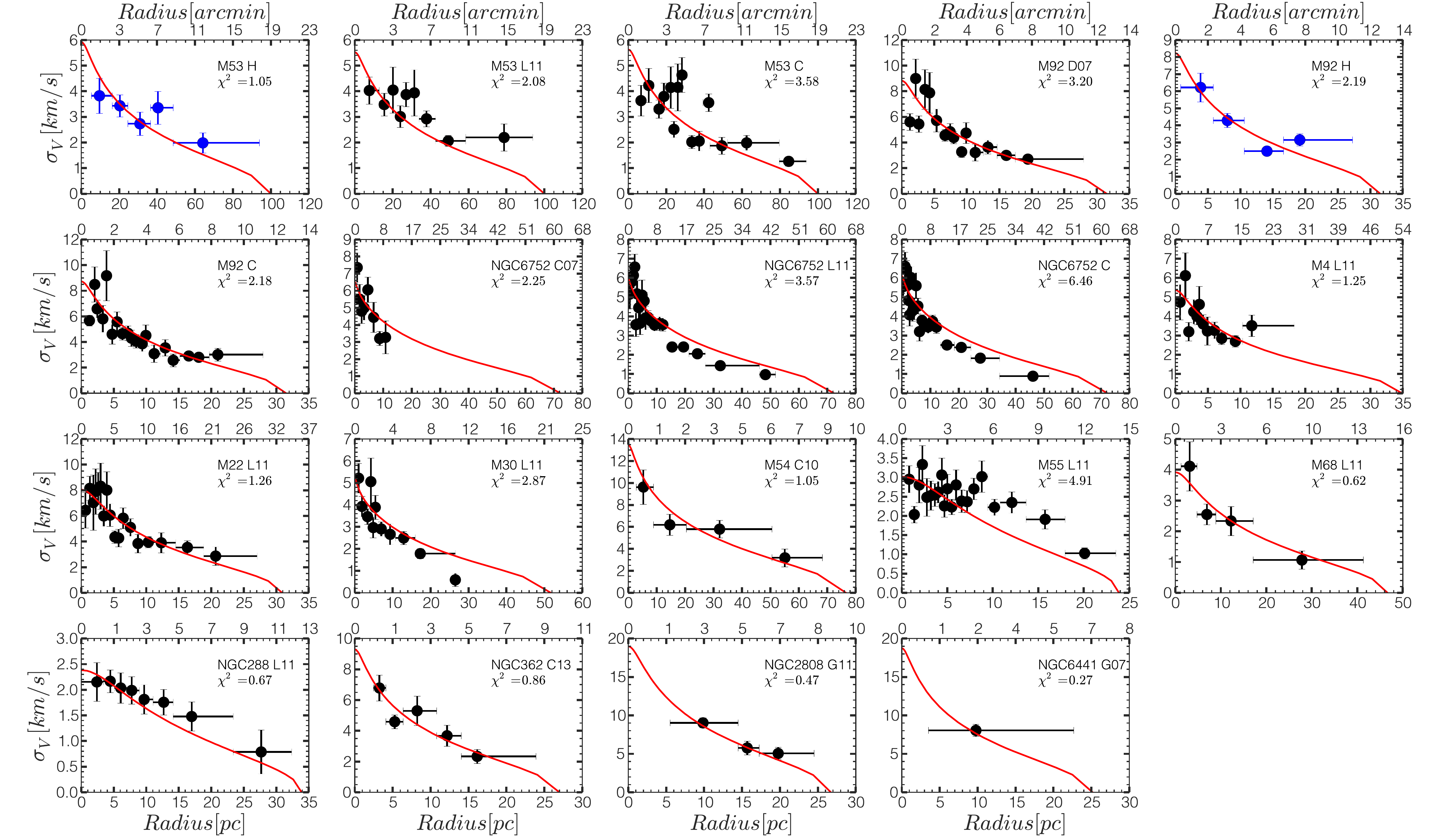}
\caption{The dispersion profile versus plotted for all clusters. Plots appear in the same order as Table \ref{tab:DispMass}. Data points represent the binned dispersion profile, Hectochelle data are shown in blue. The red line shows the best fit K66 profile. In the upper right we label the cluster and data source (from Table \ref{tab:DispMass}) as well as the reduced $\chi^2$ value. Binned dispersions are determined for stars with memberships $> 0.5$, and the maximum number of bins is limited to 20.}
\label{fig:multi_king}
\end{figure*}

In Figure \ref{fig:m92radialdisp} we show the binned dispersions plotted over their K66 fit for M~92. The data is binned radially in bins of 20 stars. There is good agreement between the model and data. Figure \ref{fig:m92radialdisp} also shows our Hectochelle data is in very good agreement with the data published in \cite{drukier2007}.  The binned dispersions and best-fits models for all the datasets are shown in Figure \ref{fig:multi_king}. We determine binned dispersions for stars with membership probability $\geq 0.5$, and set the maximum number of bins to be 20. To help assess the goodness of fit of the King models, we calculate a reduced $\chi^2$ by comparing the model predictions to the binned dispersions and assuming no degrees of freedom.  These $\chi^2$ values are shown in Table \ref{tab:DispMass} and listed in Figure~5 in each panel. The King models are excellent fits in most cases.  We discuss the $\chi^2$ values further in section 4.3 where we assess the appropriateness of our King models.

To obtain the mass of the cluster we scale the K66 model mass, created from the core radius and concentration parameter in \cite{harris1996} (2010 edition) and a dispersion of 1 km/s, by the central dispersion obtained from the K66 models best fit to the data.
 
\subsection{Rotations}
\begin{figure}
\centering
\epsscale{1.25}
\plotone{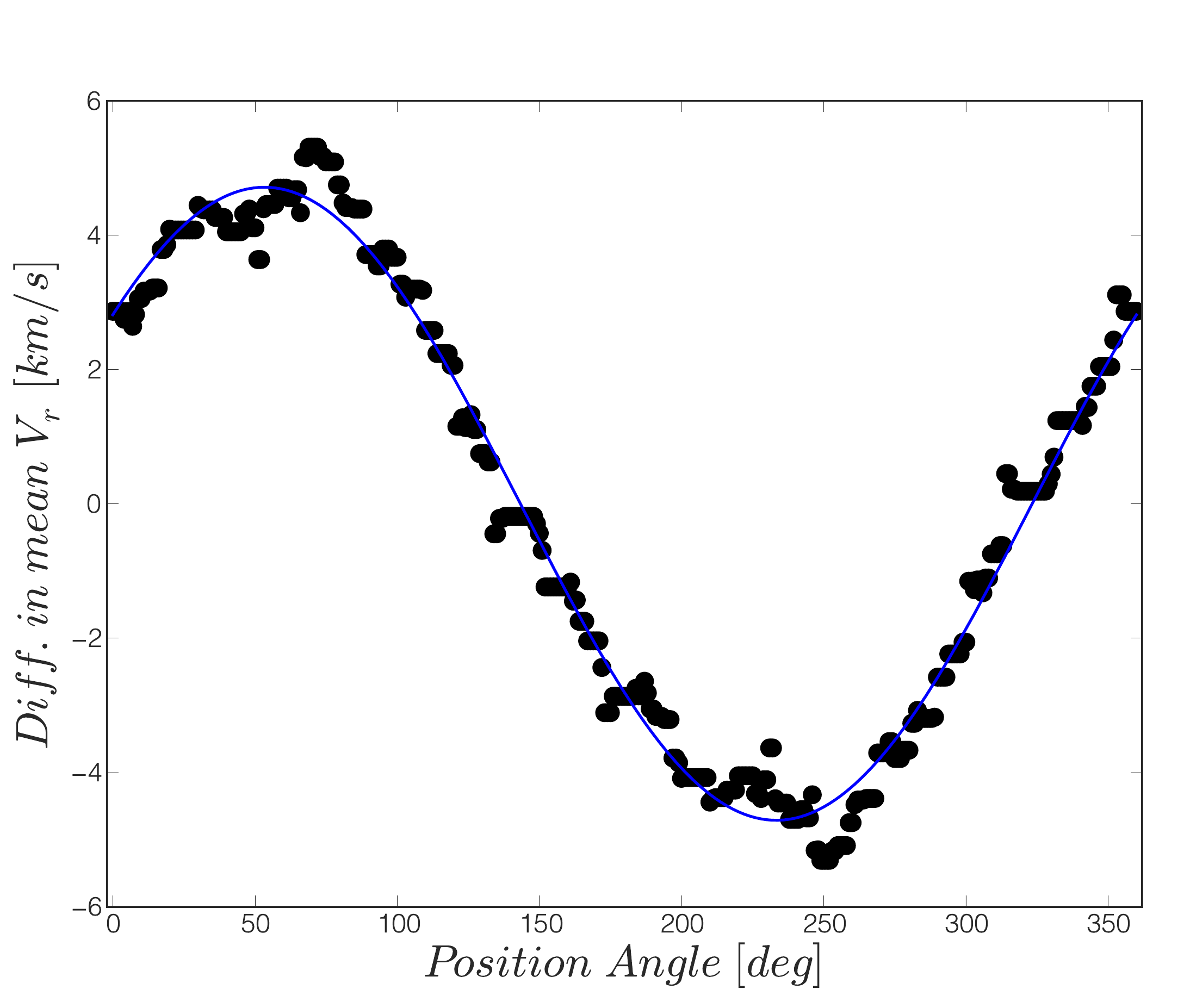}
\caption{The rotation of M~2. Plot shows the difference between the mean cluster velocity on each side a line passing through the center at angle $\phi$ measured from north to east (N=$0^\circ$ E=$90^\circ$); the position angle shown on the x-axis is defined as: $\text{PA}_0 = 270^\circ -  \phi$ and thus gives the major axis of rotation. We find a rotation of $4.7 ~\pm ~1.0$ with an approximate axis of rotation of $\text{PA}_{0}\simeq 233^{\circ}$. }
\label{fig:m2rotation}
\end{figure}

We have determined rotations for all clusters in this study using a method similar to the one described by \cite{lane2009, lane2010a, lane2010b} and \cite{bellazzini2012}. We define our sample of stars for determining rotations by selecting stars in each cluster with normalized membership probabilities $\geq$0.5. The sample is then divided in two by a line at a range of position angles (PAs) and a difference in the mean velocity of each subsample is taken.  The PA value gives the major axis of rotation in the standard convention with N at PA=$0^\circ$ and E at PA=$90^\circ$. We then fit the sine function shown by Equation \ref{eq:sine} to the mean differences and their respective PAs to obtain the maximum rotation amplitude ($A_{rot}$) and $\phi$, where $\phi$ = $270^\circ$ - $\text{PA}_0$. $\text{PA}_0$ is the position angle of the dividing line corresponding to the maximum rotation amplitude, coinciding with rotation axis. We calculated errors on the rotation via bootstrapping. As noted in \cite{bellazzini2012} $A_{rot}$ is a reasonable proxy for the actual maximum amplitude; we report $A_{rot}$ instead of $A_{rot}/2$ as in \cite{lane2010b}.  To demonstrate the method, Fig. \ref{fig:m2rotation} shows the rotation for the M~2 dataset, a cluster with a clearly detected rotation signature of $4.7 ~\pm ~1.0$ and $\text{PA}_{0}\simeq 233^{\circ}$.   Our rotation amplitude results are shown in Table \ref{tab:rotation}.  Note that not all clusters show as clear rotation as is seen in M2; this is reflected in the error bars shown in Table \ref{tab:rotation}. We compare our rotations to those found by \cite{bellazzini2012}.   We find for the sample of overlapping data that $\sim$50\% lie within 1$\sigma$ while the other $\sim$50\% lie around 2$\sigma$. We see the largest discrepancy with NGC~6441 where \cite{bellazzini2012} finds $A_{rot}$ of 12.9, almost a factor of two larger than ours. We attribute this to different membership determinations, \cite{bellazzini2012} determine rotation using all stars in the sample while we use only stars within the tidal radius with membership probabilities greater than 0.5. We also note that our errors include the effects of resampling, while those in \cite{bellazzini2012} do not, and thus our errors are typically larger.

\begin{equation}
\label{eq:sine}
\Delta \langle V_{r} \rangle = 
A_{rot}\text{sin}(\text{PA} + \phi)
\end{equation}

\section{Results and Comparison to Previous Work}
\label{sec:results}

The results of King model fits to our kinematic data are given in Table \ref{tab:DispMass} for each cluster.  The number of members in each cluster ranges from 15 to 2660, and this data yields typical errors on the $M/L_V$ and mass determinations of 15\%.  In this section we compare the results of clusters with more than one data set, and describe our combination of these data sets.  We then compare our results to previous work on determining clusters masses and mass-to-light ratios. We discuss systematic errors and modeling differences in Sec. \ref{sec:systematic_errors}.

\begin{deluxetable}{llllll}
\tablecolumns{6}
\tablewidth{0pt} 
\tabletypesize{\scriptsize}
\tablecaption{Rotational Information\label{tab:rotation}}

\tablehead{
        \colhead{Cluster} &
        \colhead{$A_{rot}$} &
        \colhead{$r_{in}/r_{h}$} &
        \colhead{$r_{out}/r_{h}$} &
        \colhead{$<r/r_{h}>$} &
        \colhead{source} 
        \\
        \colhead{} &
        \colhead{($km/s^{-1}$)} &
        \colhead{} &
        \colhead{} &
        \colhead{} &
        \colhead{} } 
\startdata
M 2 & $4.7 ~\pm ~ 1.0$ & 0.45  & 11.25  & 4.82  & H \\ 
M 3 & $0.6 ~\pm ~ 1.0$ & 0.20  & 12.29  & 4.49  & H \\ 
M 14 & $2.1 ~\pm ~ 2.0$ & 0.25  & 5.83  & 2.64  & H \\ 
M 71 & $0.4 ~\pm ~ 0.8$ & 0.33  & 5.25  & 3.06  & H \\ 
NGC 4147 & $1.6 ~\pm ~ 0.9$ & 1.37  & 9.74  & 4.40  & H \\ 
NGC 5053 & $0.3 ~\pm ~ 0.5$ & 0.14  & 3.88  & 1.64  & H \\ 
NGC 5466 & $1.1 ~\pm ~ 0.3$ & 0.10  & 6.28  & 2.25  & H \\ 
NGC 6934 & $1.8 ~\pm ~ 1.4$ & 0.72  & 10.74  & 5.70  & H \\ 

47 Tuc & $4.7 ~\pm~ 5.1$ & 0.34 & 3.92 & 1.49 & G13 \\ 
\textbf{47 Tuc} & $\bm{4.0 ~\pm ~ 0.3}$ & \textbf{0.07}  & \textbf{13.34}  & \textbf{4.26}  & \textbf{L11} \\ 
47 Tuc & $4.0~\pm~0.3$ & 0.07 & 13.34 & 4.17 & C \\

M 5 & $2.2~\pm~0.9$ & 0.55 & 6.98 & 2.45 & G13 \\ 
\textbf{M 5} & $\bm{2.1 ~\pm ~ 0.7}$ & \textbf{0.70}  & \textbf{13.14}  & \textbf{5.46}  & \textbf{H} \\ 
M 5 & $2.1~\pm~0.6$ & 0.55 & 13.14 & 4.36 & C \\

M 12 & $0.3~\pm~0.6$ & 0.07 & 9.74 & 3.51 & L11 \\ 
M 12 & $1.0~\pm~0.7$ & 0.93 & 5.34 & 2.36 & C07b \\ 
\textbf{M 12} & $\bm{0.2 ~\pm ~ 0.5}$ & \textbf{0.07}  & \textbf{9.74}  & \textbf{3.30}  & \textbf{C} \\ 

M 15 & $2.5~\pm~0.9$ & 0.13 & 16.61 & 4.36 & D98 \\ 
M 15 & $2.5~\pm~2.8$ & 0.53 & 17.15 & 6.18 & H \\ 
\textbf{M 15} & $\bm{2.5 ~\pm ~ 0.8}$ & \textbf{0.13}  & \textbf{17.15}  & \textbf{4.85}  & \textbf{C} \\ 

M 53 & $0.3~\pm~0.8$ & 0.77 & 13.76 & 5.48 & H \\ 
M 53 & $0.7~\pm~0.8$ & 0.85 & 13.96 & 5.41 & L11 \\ 
\textbf{M 53} & $\bm{0.4 ~\pm ~ 0.7}$ & \textbf{0.77}  & \textbf{13.96}  & \textbf{5.43}  & \textbf{C} \\

M 92 & $2.1~\pm~0.9$ & 0.23 & 11.35 & 3.29 & D07 \\ 
M 92 & $0.9~\pm~1.6$ & 0.32 & 11.07 & 4.98 & H \\ 
\textbf{M 92} & $\bm{1.8 ~\pm ~ 0.8}$ & \textbf{0.23}  & \textbf{11.35}  & \textbf{3.67}  & \textbf{C} \\ 

NGC 6752 & $1.2~\pm~1.2$ & 0.14 & 5.41 & 2.02 & C07 \\ 
NGC 6752 & $0.4~\pm~0.5$ & 0.03 & 28.13 & 15.57 & L11 \\ 
\textbf{NGC 6752} & $\bm{0.3 ~\pm ~ 0.5}$ & \textbf{0.03}  & \textbf{28.13}  & \textbf{14.88}  & \textbf{C} \\

M 4 & $1.3 ~\pm ~ 0.5$ & 0.04  & 11.95  & 4.89  & L11 \\ 
M 22 & $2.5 ~\pm ~ 2.3$ & 0.06  & 9.48  & 5.06  & L11 \\ 
M 30 & $0.5 ~\pm ~ 0.8$ & 0.06  & 18.32  & 7.96  & L11 \\ 
M 54 & $1.6 ~\pm ~ 2.9$ & 0.44  & 11.43  & 4.09  & C10 \\ 
M 55 & $0.4 ~\pm ~ 0.2$ & 0.01  & 5.41  & 2.15  & L11 \\ 
M 68 & $0.4 ~\pm ~ 1.0$ & 0.27  & 9.82  & 4.54  & L11 \\ 
NGC 288 & $0.9 ~\pm ~ 0.8$ & 0.01  & 5.90  & 2.17  & L11 \\ 
NGC 362 & $1.6 ~\pm ~ 1.4$ & 1.19  & 11.65  & 4.13  & C13 \\ 
NGC 2808 & $3.1 ~\pm ~ 3.8$ & 2.47  & 11.02  & 6.98  & G11 \\ 
NGC 6441 & $7.3 ~\pm ~ 4.5$ & 1.83  & 12.08  & 6.89  & G07  
\enddata
\tablenotetext{Note:}{Columns from left to right. (1) Cluster name, (2) rotation determined by method described in \cite{bellazzini2012}, (3) innermost radii of star over the effective radius, (4) outermost radii of star over the effective radius, (5) the average of the radii over the effective radius, (6) source - H stands for Hectochelle, C stands for combined, and all other sources can be found in Table \ref{tab:archival}.  In combined datasets section, bold face indicates the best fit for each cluster.}
\end{deluxetable}

\begin{deluxetable*}{llllllllll}
\tablecolumns{10}
\tablewidth{0pt} 
\tabletypesize{\scriptsize}
\tablecaption{Velocity and Dispersion Information \label{tab:DispMass}}

\tablehead{
    \colhead{Cluster} & 
    \colhead{[Fe/H]} & 
    \colhead{$\overline{V}$} & 
    \colhead{$\sigma_{0}$} & 
    \colhead{$M/L_{V}$} & 
    \colhead{$\log M_{King}$} & 
    \colhead{$\chi^2$} & 
    \colhead{$N_{tidal}$} & 
    \colhead{$N_{mem}$} & 
    \colhead{Source}
    \\ 
    \colhead{} & 
    \colhead{(dex)} & 
    \colhead{($\text{km} \ \text{s}^{-1}$)} & 
    \colhead{($\text{km} \ \text{s}^{-1}$)} & 
    \colhead{($M_\odot/L_\odot$)} & 
    \colhead{($M_\odot$)} & 
    \colhead{} & 
    \colhead{} &
    \colhead{} &
    \colhead{}} 

\startdata
\multicolumn{10}{c}{\bf Hectochelle Data} \\
\cline{1-10} \\
M 2 & -1.65 & $-2.2 ~\pm ~ 0.7$ & $10.4 ~\pm ~ 1.2$ & $1.66 ~\pm ~ 0.38$ & $5.76 ~\pm ~ 0.20$ & 0.68 & 91 & 80 & H \\ 
M 3 & -1.50 & $-147.4 ~\pm ~ 0.3$ & $7.4 ~\pm ~ 0.6$ & $1.53 ~\pm ~ 0.23$ & $5.67 ~\pm ~ 0.13$ & 1.12 & 157 & 139 & H \\ 
M 14 & -1.28 & $-60.0 ~\pm ~ 0.9$ & $11.7 ~\pm ~ 1.1$ & $1.82 ~\pm ~ 0.35$ & $5.83 ~\pm ~ 0.17$ & 0.46 & 103 & 91 & H \\ 
M 71 & -0.78 & $-23.1 ~\pm ~ 0.3$ & $3.1 ~\pm ~ 0.5$ & $1.36 ~\pm ~ 0.39$ & $4.31 ~\pm ~ 0.25$ & 0.33 & 91 & 43 & H \\ 
NGC 4147 & -1.80 & $+179.5 ~\pm ~ 0.5$ & $3.2 ~\pm ~ 0.6$ & $1.47 ~\pm ~ 0.54$ & $4.57 ~\pm ~ 0.33$ & 0.51 & 15 & 15 & H \\ 
NGC 5053 & -2.27 & $+42.6 ~\pm ~ 0.3$ & $1.8 ~\pm ~ 0.2$ & $1.3 ~\pm ~ 0.26$ & $4.75 ~\pm ~ 0.18$ & 0.55 & 50 & 38 & H \\ 
NGC 5466 & -1.98 & $+106.9 ~\pm ~ 0.2$ & $1.5 ~\pm ~ 0.2$ & $0.72 ~\pm ~ 0.23$ & $4.58 ~\pm ~ 0.28$ & 1.56 & 51 & 47 & H \\ 
NGC 6934 & -1.47 & $-406.5 ~\pm ~ 0.5$ & $5.2 ~\pm ~ 0.8$ & $1.52 ~\pm ~ 0.49$ & $5.10 ~\pm ~ 0.28$ & 1.97 & 48 & 24 & H \\ 

\cutinhead{\textbf{Combined Datasets}}
47 Tuc & -0.72 & $-19.6 ~\pm ~ 0.8$ & $13.2 ~\pm ~ 0.8$ & $1.76 ~\pm ~ 0.22$ & $5.95 ~\pm ~ 0.11$ & 6.41 & 110 & 110 & G13 \\ 
\textbf{47 Tuc} & \textbf{-0.72} & $\bm{-16.8 ~\pm ~ 0.2}$ & $\bm{11.3 ~\pm ~ 0.2}$ & $\bm{1.30 ~\pm ~ 0.05}$ & $\bm{5.81 ~\pm ~ 0.03}$ & \textbf{3.16} & \textbf{3195} & \textbf{2548} &\textbf{L11} \\ 
47 Tuc & -0.72 & $-16.9 ~\pm ~ 0.2$ & $11.5 ~\pm ~ 0.2$ & $1.35 ~\pm ~ 0.05$ & $5.83 ~\pm ~ 0.03$ & 3.33 & 3305 & 2660 & C \\ 
M 5 & -1.29 & $+52.5 ~\pm ~ 0.4$ & $5.7 ~\pm ~ 0.6$ & $0.67 ~\pm ~ 0.14$ & $5.28 ~\pm ~ 0.18$ & 1.26 & 90 & 90 & G13 \\ 
\textbf{M 5} & \textbf{-1.29} & $\bm{+53.9 ~\pm ~ 0.4}$ & $\bm{8.1 ~\pm ~ 0.6}$ & $\bm{1.36 ~\pm ~ 0.21}$ & $\bm{5.59 ~\pm ~ 0.13}$ & \textbf{0.82} & \textbf{155} & \textbf{128} & \textbf{H} \\ 
M 5 & -1.29 & $+53.3 ~\pm ~ 0.3$ & $7.0 ~\pm ~ 0.5$ & $1.01 ~\pm ~ 0.14$ & $5.46 ~\pm ~ 0.12$ & 1.60 & 245 & 217 & C \\ 
M 12 & -1.37 & $-41.0 ~\pm ~ 0.3$ & $4.0 ~\pm ~ 0.4$ & $0.89 ~\pm ~ 0.16$ & $4.81 ~\pm ~ 0.15$ & 1.81 & 407 & 285 & L11 \\ 
M 12 & -1.37 & $-42.9 ~\pm ~ 0.4$ & $4.4 ~\pm ~ 0.3$ & $1.07 ~\pm ~ 0.16$ & $4.89 ~\pm ~ 0.13$ & 0.89 & 92 & 81 &C07b \\
\textbf{M 12} & \textbf{-1.37} & $\bm{-41.5 ~\pm ~ 0.2}$ & $\bm{4.3 ~\pm ~ 0.2}$ & $\bm{1.02 ~\pm ~ 0.11}$ & $\bm{4.87 ~\pm ~ 0.10}$ & \textbf{2.05} & \textbf{499} & \textbf{369} & \textbf{C} \\ 
M 15 & -2.37 & $-106.7 ~\pm ~ 0.4$ & $8.2 ~\pm ~ 0.4$ & $1.29 ~\pm ~ 0.14$ & $5.72 ~\pm ~ 0.09$ & 7.08 & 230 & 230 & D98 \\ 
M 15 & -2.37 & $-107.4 ~\pm ~ 0.6$ & $7.7 ~\pm ~ 1.0$ & $1.13 ~\pm ~ 0.29$ & $5.66 ~\pm ~ 0.22$ & 2.65 & 84 & 73 & H \\ 
\textbf{M 15} & \textbf{-2.37} & $\bm{-106.9 ~\pm ~ 0.3}$ & $\bm{8.1 ~\pm ~ 0.4}$ & $\bm{1.26 ~\pm ~ 0.13}$ & $\bm{5.71 ~\pm ~ 0.09}$ & \textbf{19.30} & \textbf{314} & \textbf{303} &\textbf{C} \\ 
M 53 & -2.10 & $-62.7 ~\pm ~ 0.3$ & $5.9 ~\pm ~ 0.6$ & $1.48 ~\pm ~ 0.31$ & $5.59 ~\pm ~ 0.18$ & 1.05 & 110 & 101 & H \\ 
M 53 & -2.10 & $-62.9 ~\pm ~ 0.3$ & $5.5 ~\pm ~ 0.6$ & $1.28 ~\pm ~ 0.26$ & $5.52 ~\pm ~ 0.18$ & 2.08 & 216 & 169& L11 \\ 
\textbf{M 53} & \textbf{-2.10} & $\bm{-62.8 ~\pm ~ 0.3}$ & $\bm{5.6 ~\pm ~ 0.4}$ & $\bm{1.33 ~\pm ~ 0.20}$ & $\bm{5.54 ~\pm ~ 0.13}$ & \textbf{3.58} & \textbf{326} & \textbf{269} &\textbf{C} \\ 
M 92 & -2.31 & $-121.0 ~\pm ~ 0.3$ & $8.8 ~\pm ~ 0.5$ & $1.70 ~\pm ~ 0.18$ & $5.45 ~\pm ~ 0.09$ & 3.20 & 300 & 299 & D07 \\ 
M 92 & -2.31 & $-121.4 ~\pm ~ 0.5$ & $8.2 ~\pm ~ 0.6$ & $1.48 ~\pm ~ 0.23$ & $5.39 ~\pm ~ 0.13$ & 2.19 & 85 & 77 & H \\ 
\textbf{M 92} & \textbf{-2.31} & $\bm{-121.1 ~\pm ~ 0.3}$ & $\bm{8.7 ~\pm ~ 0.5}$ & $\bm{1.67 ~\pm ~ 0.17}$ & $\bm{5.44 ~\pm ~ 0.09}$ & \textbf{2.18} & \textbf{385} & \textbf{376} & \textbf{C} \\ 
NGC 6752 & -1.54 & $-26.1 ~\pm ~ 0.5$ & $6.5 ~\pm ~ 0.5$ & $3.06 ~\pm ~ 0.44$ & $5.51 ~\pm ~ 0.12$ & 2.25 & 151 & 148 & C07 \\ 
NGC 6752 & -1.54 & $-26.1 ~\pm ~ 0.2$ & $6.0 ~\pm ~ 0.3$ & $2.55 ~\pm ~ 0.22$ & $5.43 ~\pm ~ 0.07$ & 3.57 & 2809 & 630 &L11 \\ 
\textbf{NGC 6752} & \textbf{-1.54} & $\bm{-26.1 ~\pm ~ 0.2}$ & $\bm{6.1 ~\pm ~ 0.2}$ & $\bm{2.69 ~\pm ~ 0.20}$ & $\bm{5.45 ~\pm ~ 0.06}$ & \textbf{6.46} & \textbf{2960} & \textbf{779} & \textbf{C} \\ 
\cutinhead{\textbf{Archival Data}}
M 4 & -1.16 & $+71.6 ~\pm ~ 0.3$ & $5.4 ~\pm ~ 0.4$ & $1.81 ~\pm ~ 0.25$ & $5.07 ~\pm ~ 0.12$ & 1.25 & 540 & 234 & L11 \\ 
M 22 & -1.70 & $-145.2 ~\pm ~ 0.4$ & $8.1 ~\pm ~ 0.6$ & $1.43 ~\pm ~ 0.23$ & $5.49 ~\pm ~ 0.14$ & 1.26 & 1196 & 349 &L11 \\ 
M 30 & -2.27 & $-183.8 ~\pm ~ 0.4$ & $5.2 ~\pm ~ 0.3$ & $1.76 ~\pm ~ 0.22$ & $5.16 ~\pm ~ 0.11$ & 2.87 & 427 & 203 & L11 \\ 
M 54 & -1.49 & $+143.5 ~\pm ~ 1.1$ & $13.5 ~\pm ~ 2.0$ & $1.52 ~\pm ~ 0.45$ & $6.11 ~\pm ~ 0.26$ & 1.05 & 75 & 66 &C10 \\ 
M 55 & -1.94 & $+177.3 ~\pm ~ 0.1$ & $3.0 ~\pm ~ 0.1$ & $0.61 ~\pm ~ 0.06$ & $4.74 ~\pm ~ 0.08$ & 4.91 & 1190 & 773 & L11 \\ 
M 68 & -2.23 & $-95.0 ~\pm ~ 0.4$ & $3.9 ~\pm ~ 0.6$ & $1.4 ~\pm ~ 0.43$ & $5.03 ~\pm ~ 0.27$ & 0.62 & 135 & 77 & L11 \\ 
NGC 288 & -1.32 & $-45.2 ~\pm ~ 0.2$ & $2.4 ~\pm ~ 0.3$ & $1.08 ~\pm ~ 0.27$ & $4.66 ~\pm ~ 0.22$ & 0.67 & 192 & 146 & L11 \\ 
NGC 362 & -1.26 & $+223.3 ~\pm ~ 0.5$ & $9.3 ~\pm ~ 0.7$ & $1.24 ~\pm ~ 0.19$ & $5.40 ~\pm ~ 0.13$ & 0.86 & 91 & 91 & C13 \\ 
NGC 2808 & -1.14 & $+102.2 ~\pm ~ 1.1$ & $19.0 ~\pm ~ 1.9$ & $2.5 ~\pm ~ 0.51$ & $6.09 ~\pm ~ 0.18$ & 0.47 & 70 & 53 & G11 \\ 
NGC 6441 & -0.46 & $+19.0 ~\pm ~ 3.0$ & $18.7 ~\pm ~ 5.6$ & $1.58 ~\pm ~ 0.94$ & $5.98 ~\pm ~ 0.53$ & 0.27 & 26 & 19 & G07  
\enddata
\tablenotetext{Note:}{Columns from left to right. (1) Cluster name, (2) metallicity reported by \cite{harris1996} (2010 edition), (3) average velocity, (4) central dispersion, (5) mass to light ratio, (6) log of the K66 mass in solar units, (7) chi squared value for the fit, (8) number of stars within the tidal radius, (9) number of members stars ($P_M \geq 0.5$), (10) source - H stands for Hectochelle, C stands for combined, and all other sources can be found in Table \ref{tab:archival}.  In combined datasets section, bold face indicates the best fit for each cluster.}
\end{deluxetable*}

\subsection{Clusters with Multiple Data Sets}

For seven of our Hectochelle clusters, at least one additional data set is available; multiple data sets also exist for three clusters without Hectochelle data. We start by comparing these data sets to each other to ensure that they are consistent.  Where consistency is found, we combine the data sets to provide the best possible kinematic sample.  Fig.~\ref{fig:m92radialdisp} shows the comparison of the best fits to K66 models for M~92.  This is a case where good consistency is seen in both the best fit velocity and central velocity dispersion between the Hectochelle and \cite{drukier2007} data sets.  The agreement is at the 1$\sigma$ level in both quantities.  Good agreement is also seen in M~12, M~15, and M~53; for each of these we fit combined data sets to obtain our best parameter estimates; these are shown in bold in Table \ref{tab:DispMass}.  

However, for 47~Tuc and M~5, the agreement between data sets is less good.  In 47~Tuc, the mean velocity estimates between the \cite{carretta2013a} and \cite{lane2011} are almost 3~km/s apart, $>$3$\sigma$.  In M~5, both the velocity and dispersion measurements are inconsistent between the Hectochelle and \cite{carretta2013a} data at the $\sim$3$\sigma$ level.  Both cases involve the \cite{carretta2013a} data, where errors are not given for each individual velocity.  Given the importance of correct error estimates for dispersion determinations, we assume these data sets are less reliable, and thus we prefer the \cite{lane2011} data for 47~Tuc and the Hectochelle data for M~5.  These preferred data sets are shown in bold in Table \ref{tab:DispMass}. 

\subsection{Comparison with Previous Work}

Central dispersion estimates have previously been obtained for all clusters in our sample apart from M~14.  In 2 cases (NGC~6441 and NGC~2808), these previous estimates are single integrated central dispersion measurements. For the remaining clusters, they result from modeling of individual stellar velocity data similar to that presented here, but often with different modeling assumptions.  Here we compare our derived masses to these previous mass estimates and discuss differences in modeling approaches.  

We first describe the measurements and modeling methods of the studies to which we compare our data.  The kinematics are either integrated measurements \citep{dubath1997} or individual stellar velocities \citep{lane2010b,bellazzini2012}, or both \citep{pryor1993,mandushev1991,mclaughlin2005}.  These measurements were modeled to derive central velocity dispersions, mass-to-light ratios and masses.  Multi-mass king models were used by \cite{pryor1993}, while single mass King models were used by \cite{mandushev1991}, \cite{mclaughlin2005}, and \cite{bellazzini2012}.  Plummer model fits using a Markov Chain Monte Carlo method were performed by \cite{lane2010b} to derive velocity dispersions and masses; these models don't incorporate surface brightness profile information as the authors were interested in deriving radial $M/L$ variations.  Thus our comparisons of all quantities are to somewhat heterogeneous data; we expect to see variations between our findings and previous work due to the data quality, sample size and methods.

\begin{figure}
\centering
\epsscale{1.25}
\plotone{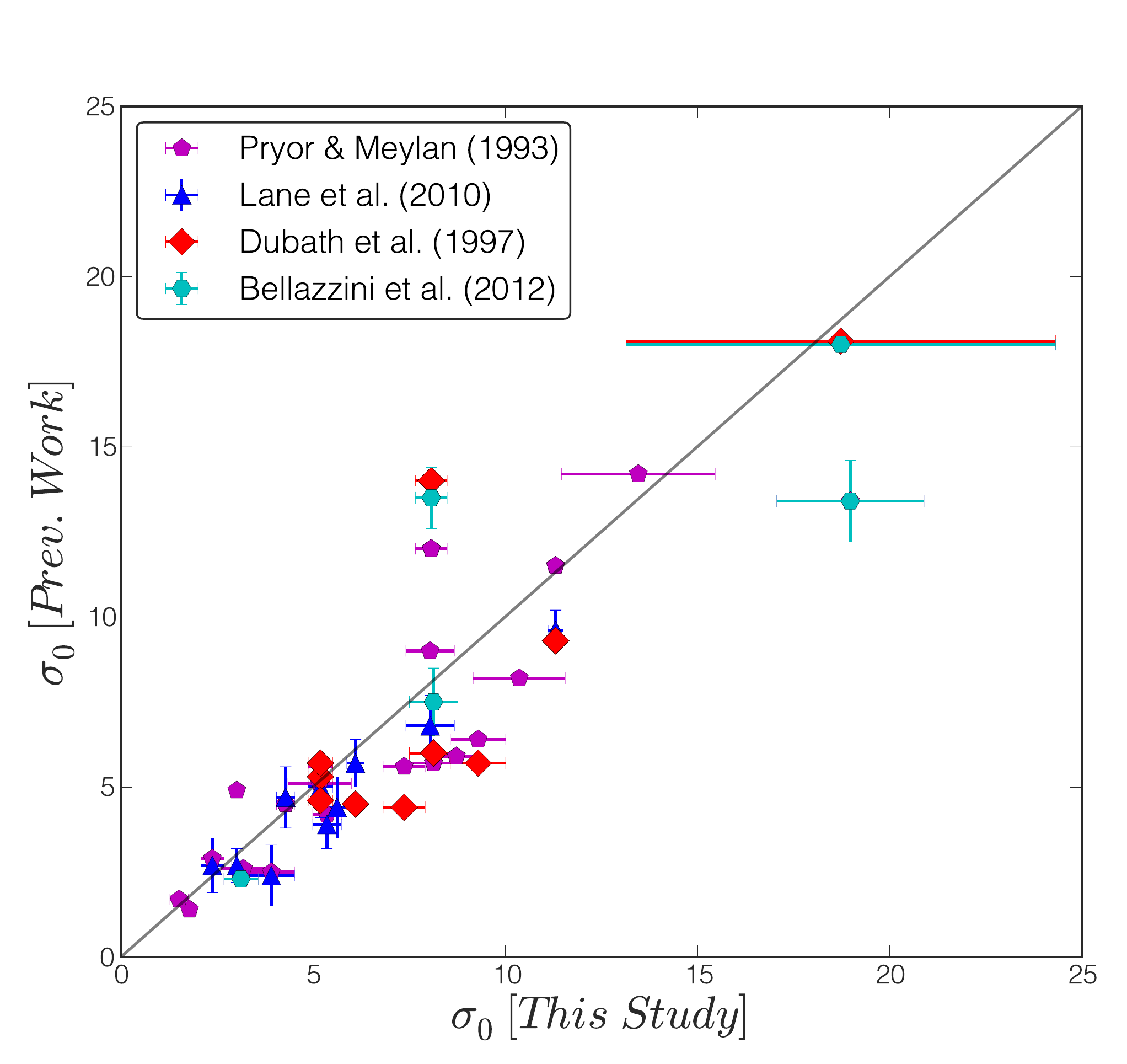}
\caption{Comparison of the central velocity dispersions found in this study 
against those of previous work.}\label{fig:dispCompare}
\end{figure}

Figure \ref{fig:dispCompare} shows the central dispersions derived in this study against those derived in previous work. Overall, the dispersion estimates agree quite well, within 1$\sigma$ for $\sim$50\% of the \cite{lane2010b} and \cite{bellazzini2012} values. We see the largest discrepancies between {bf our data and the work of} \cite{pryor1993} and \cite{dubath1997} where only $\sim$30\% and $\sim$12\% of the values compared are within 1$\sigma$. There does appear to be a slight bias towards higher central dispersion estimates in our data, especially when comparing the multi-mass King model fits of \cite{pryor1993} and the integrated central dispersions of \cite{dubath1997}. We discuss these differences in the next subsection.

\begin{figure}
\centering
\epsscale{1.25}
\plotone{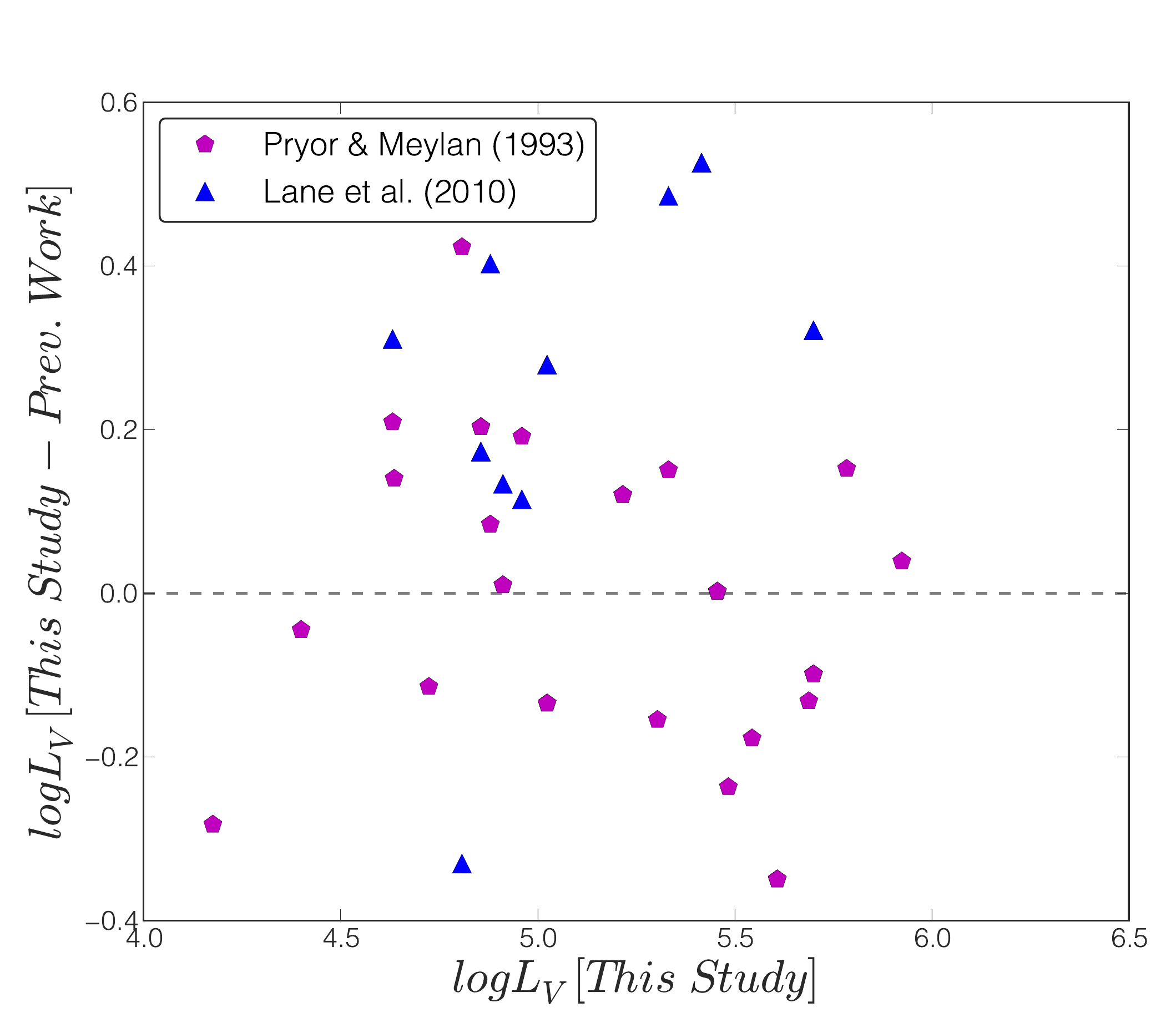}
\caption{The difference between luminosities found in this study and previous 
studies.}\label{fig:luminositycompare}
\end{figure}

In Figure \ref{fig:luminositycompare} we show the difference between our luminosity and the luminosities found by \cite{pryor1993} and \cite{lane2010b}. We see large discrepancies.  With the goal of trying to use the most accurate parameters, we have chosen to use the  \cite{harris1996} (2010 edition) catalogue as our source for luminosities.  The luminosities are dependent on horizontal branch distance estimates from a variety of sources; of these two thirds have been updated since \cite{pryor1993}.  On the other hand, the luminosities in \cite{lane2010b} come from their own \cite{plummer1911} model fits to surface brightness profiles. These improved distance estimates suggest that the luminosity data represents an improvement in data quality and uniformity over previous large compilations of mass-to-light ratios by \cite{pryor1993} and \cite{mandushev1991}. 

\begin{figure}
\centering
\epsscale{1.25}
\plotone{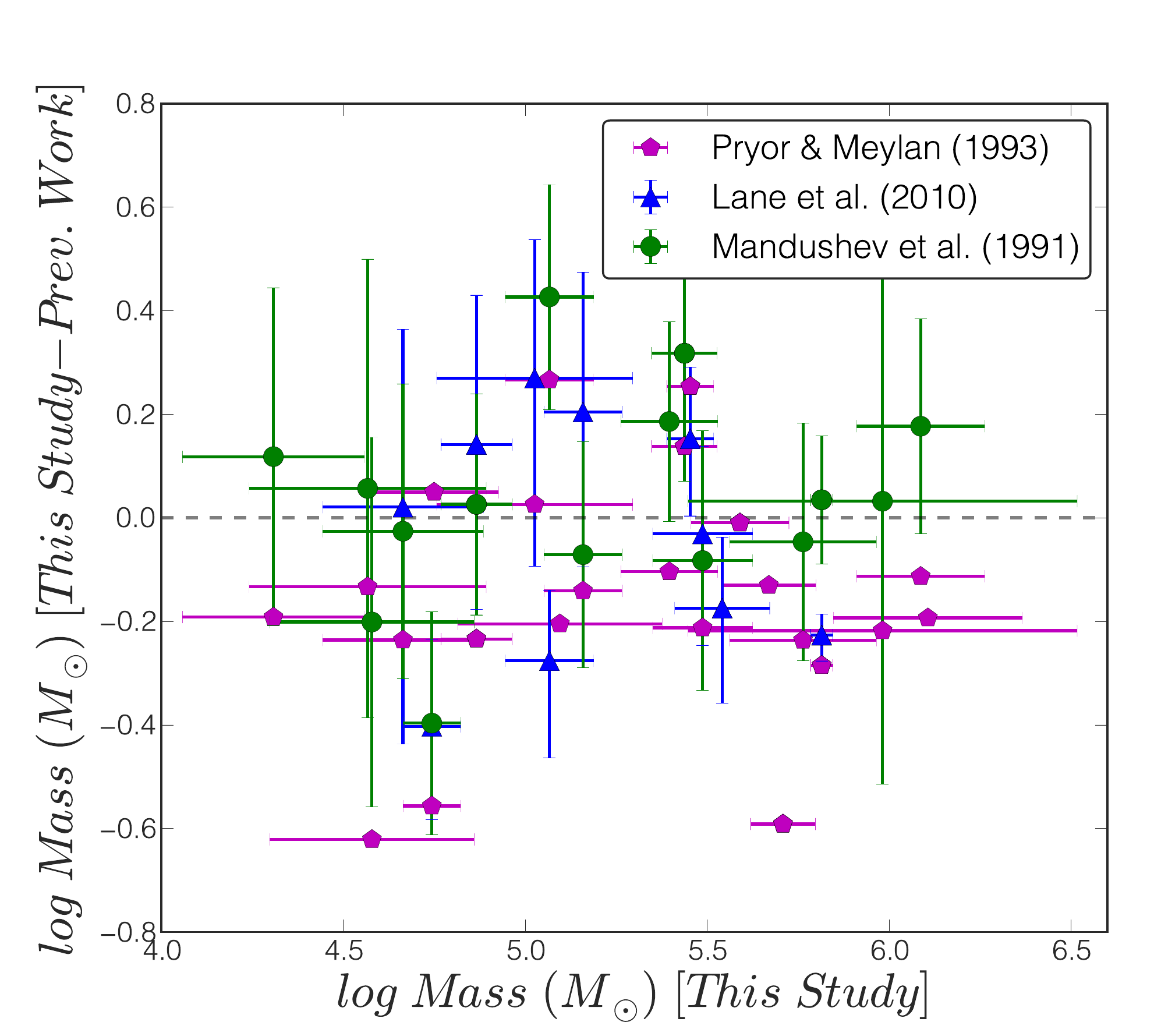}
\caption{Comparison of masses derived in this study against those 
of previous work.}\label{fig:massCompare}
\end{figure}

A comparison of our derived masses to those in previous work are shown in Figure~\ref{fig:massCompare}.  As with the dispersions, the masses are broadly consistent. The agreement is best with the \cite{lane2010b} masses and single mass King model estimates of \cite{mandushev1991} showing agreement on the 1$\sigma$ level for $\sim$70\% of the clusters compared. We don't expect our masses to differ greatly with \cite{lane2010b} as we are using the same kinematic data in our mass determinations. Our mass estimates are systematically $\sim$36\% lower than the multi-mass King model estimates of \cite{pryor1993}.  We discuss this discrepancy in the next subsection.

\begin{figure}
\centering
\epsscale{1.25}
\plotone{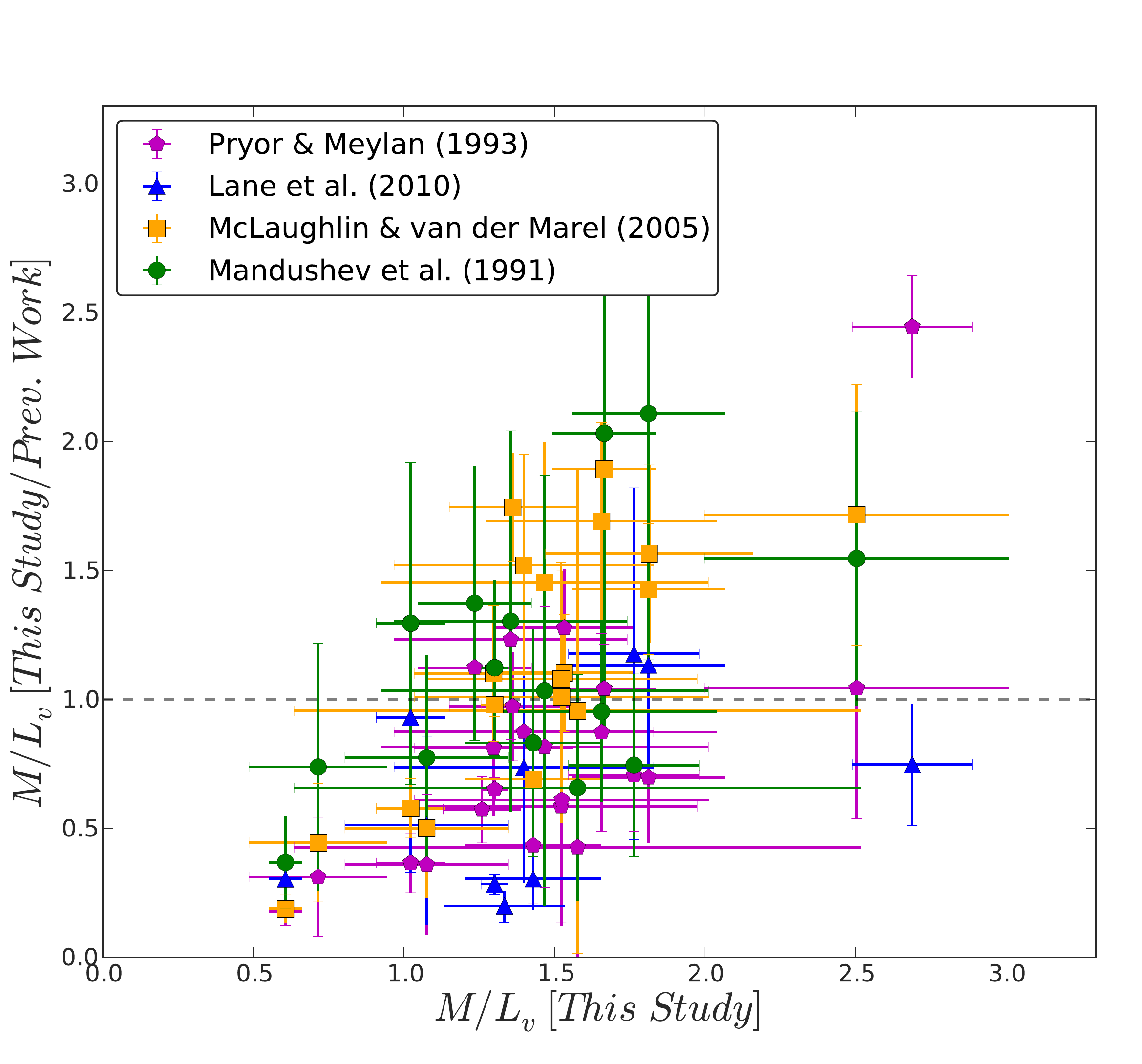}
\caption{The ratio of $M/L$s found in this study with $M/L$s 
of previous work.}\label{fig:mtlcompare}
\end{figure}

Finally, Figure \ref{fig:mtlcompare} compares the $M/L_V$s derived in this study to previous work. In general, we find good agreement ($\sim$ 80\% within 1$\sigma$) for the majority of the clusters when looking at \cite{mandushev1991} and \cite{mclaughlin2005}, who both use K66 models, while the \cite{pryor1993} values are typically 40\% higher.  We see large disagreements with the $M/L$ values of \cite{lane2010b}.  Given the general agreement of masses, this is likely due to differences in the luminosity estimates.  The largest discrepancy is for M~53, where they derive a $M/L_v$ of 6.7, while our $M/L_V$ is just 1.33$~\pm~$0.20. The independent determination of the surface brightness and mass profile used by \cite{lane2009} makes their approach significantly different from all other studies. There are numerous differences in the way $M/L$ values are derived in their work compared to ours. For M~53, \cite{lane2009} are not able to obtain an $M/L$ for the central region resulting in a large difference from our value. We note that for NGC~6752, the core collapsed profile may make the K66 model inappropriate for estimating the mass (discussed further in the next section).  However, we note relatively good agreement between their $M/L$=3.6$~\pm~$1.1 \citep{lane2010b} and our $M/L$=2.69$~\pm~$0.20.   

\subsection{Systematic Errors and the Validity of Modeling Assumptions}
\label{sec:systematic_errors}

We model our kinematic data with K66 models to facilitate comparison between Galactic and extragalactic globular cluster $M/L$ estimates \citep[e.g.][]{strader2011}. In this section we first examine how well the K66 models match the data, estimate the systematic errors in our mass determinations, and then discuss the differences between our modeling approach and other previous work. 

The K66 models provide remarkably good fits to our kinematic data.  Fig.~\ref{fig:multi_king} shows the best fit King models for each data set and the agreement between these and the binned dispersions is very good for most clusters. For 18/25 clusters fit, the reduced $\chi^2$ values are less than 2.  In fact, for the Hectochelle data where we have verified our velocity errors with repeat measurements (Fig.~2), the highest $\chi^2$ value is 2.19 in M92.  Thus King models seem to be quite good fits to these data sets.  We discuss the two clusters with the poorest fits (M~15 and NGC~6752) below.  We note that none of the errors in the archival data are verified with repeat measurements, and thus the high $\chi^2$ values may be in part due to an underestimation of the velocity errors in this archival data. 

\begin{figure}
\centering
\epsscale{1.25}
\plotone{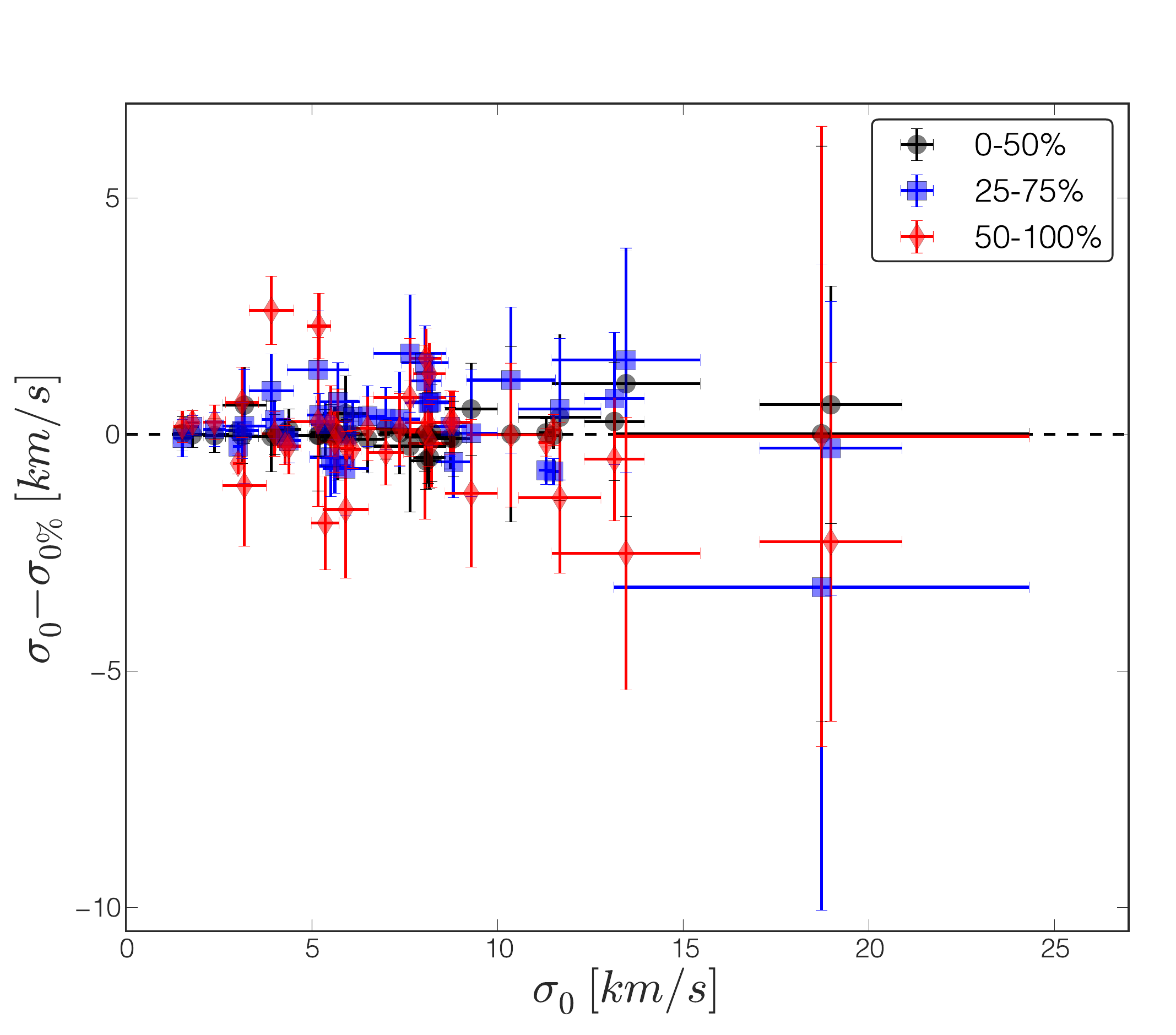}
\caption{A plot of the difference between the central dispersion derived with  all data out to the tidal radius and the central dispersion derived with percentages of the radially sorted data. Black points represent the first 50\%, blue represents 25-50\%, and red points represent 50-100\% of the radially sorted data.}\label{fig:percentDisp}
\end{figure}

To quantify how deviations from a K66 profile may affect our mass estimates we experimented with fitting restricted radial subsets of our data.  Specifically, in each cluster we fit half the stellar velocities to K66 profiles in three radial bins, an inner bin (the inner 50\%ile of radii), a middle bin (25\%ile to 75\%ile) and an outer bin (50\%ile to 100\%ile).  This approach tests for several possible systematic errors: (1) mismatches of the data to the K66 profile, (2) observational effects such as blending of stars near the center, and (3) the possible enhancement of dispersion due to binary stars in the outer parts of the cluster.  Figure \ref{fig:percentDisp} shows the difference between the central dispersions determined from the full data set and the radially binned subsets.  In general, no large systematic errors are found, with 85\% of the data agreeing within 1$\sigma$.  Clusters which were discrepant at the $\gtrsim$2$\sigma$ level are M~15, NGC~6752 (both core collapse clusters), as well as M~68 and M~22. M~22 has a clear dip in dispersion near the center (Fig.~\ref{fig:multi_king}), this suggests possible effects of confusion. The cause of the discrepancies in M~68 is unclear. 

The two core collapse objects in our sample, M~15 and NGC~6752 show clear systematic deviations from the K66 model and have the two highest reduced $\chi^2$ values (19.3 in M~15, 6.46 in NGC~6752).  This is to be expected; core collapsed clusters show very strong variations in radial $M/L$ ratio and thus are quite different from our assumed constant $M/L$.  In M~15, detailed dynamical modeling of discrete data performed by \cite{denbrok2014} shows that the $M/L$ varies by factor of $\sim$4, with the highest values found at the center, a minimum ($M/L_V \sim 1$) at intermediate radii, and then an increase at large radii to $M/L_V \sim 2$.  These changes in $M/L$ are clearly seen from the deviations to the K66 model in Fig.~\ref{fig:multi_king}, where the dispersion is higher than the K66 model in the center, lower at intermediate radii and somewhat higher again at large radii.  The total mass found from dynamical modeling and luminosity function estimations vary from $4.4 \times 10^5$ to $5.4 \times 10^5$~M$_\odot$ \citep{pasquali2004, vandenbosch2006,denbrok2014}; our K66 model fit agrees within 1$\sigma$ of all these measurements. Thus, despite the model mismatch, the total mass estimates may be correct even in the core collapsed clusters.

Another possible source of systematic error is a misestimation of King structural parameters (e.g.~Fig.~\ref{fig:luminositycompare}).  We have used the best estimates available (\cite{harris1996} (2010 edition)). However, to understand the systematic errors due to the use of King model radii, we compared our mass estimates to those derived with an older catalogue of cluster parameters \citep{trager1993}. We found that the deviation in log mass from the masses determined with \cite{harris1996} (2010 edition) parameters was $\sim$0.05 dex, which is smaller than the typical error on the K66 mass derived from the dispersion errors alone. 

Because the K66 models are single-mass models, they ignore the effects of energy equipartition and mass segregation.  These effects are incorporated explicitly into the multi-mass models used by e.g.~\cite{pryor1993}.  \cite{mandushev1991} found that masses determined from multi-mass models can be as high as a factor of two larger than masses derived from single-mass models. This difference is due to the lower central dispersions predicted by multi-mass King models relative to single-mass King models.  This lower central dispersion results from a combination of higher mass stars at the center due to mass segregation, and slower velocities for those stars due to energy equipartition.  Thus, to fit a measured central dispersion requires a higher overall mass for a multi-mass King model than for a K66 model.  Mass segregation has been found in many globular clusters \citep{koch2004,frank2012,martinazzi2014,frank2014}. However, recent work by \cite{trenti2013} found that energy equipartition is rarely seen in GCs.  Thus both multi-mass and single-mass King models provide only approximations to real clusters. The good fits of K66 models to the dispersion profiles of most of our data suggest that our mass estimates are likely accurate and do not have significant systematic biases.  Furthermore, our mass estimates enable a direct comparison with studies of extra-galatic cluster masses and $M/L$s compare trends in our data with a much larger data set of GC mass and $M/L$s \citep[e.g.][]{strader2011}.

\section{Discussion: Trends with $M/L$ and Rotation}
\label{sec:discussion}

In this section we examine the trends of $M/L$ ratio and rotation with cluster mass and metallicity.  These trends provide insight into GC formation and evolution.

\subsection{$M/L$ vs.~Cluster Mass}

As discussed in the introduction, a positive correlation of $M/L$ increasing with increasing cluster mass has been seen from dynamical $M/L$ measurements \citep{mandushev1991,strader2011}.  \cite{kruijssen2008} provided an explanation of the increasing $M/L$ values with cluster mass, suggesting that it resulted from mass loss in the cluster.  Two body relaxation is expected to preferentially eject low mass (high $M/L$) stars from the cluster, thus lowering the clusters $M/L$ over time. Because the relaxation time increases with cluster mass, lower mass clusters have undergone more mass loss and thus have reduced $M/L$s relative to higher mass clusters.  This trend is not included in simple stellar population (SSP) models \citep[e.g.][]{bruzual2003}, and thus GCs would be expected to typically have lower $M/L$ values than predicted by the models.  In the specific models used by \cite{kruijssen2008}, the speed of dynamical evolution is quantified by a dissolution timescale which varies with cluster mass and environment.  By accounting for dynamical evolution in combination with the SSP models, \cite{kruijssen2008} found that they could explain more than 90\% of available $M/L_V$ measurements with their models.  

In follow up work on Galactic globular clusters, \cite{kruijssen2009a} found that the models are broadly consistent but slightly overpredict the observed $M/L$ values.  More recently, with observations of 163 GCs in M~31, \cite{strader2011} found that a \cite{kruijssen2008} model with a dissolution timescale of 1 Myr described the general trend seen with mass and $M/L$ very well. 

\begin{figure}
\centering
\epsscale{1.25}
\plotone{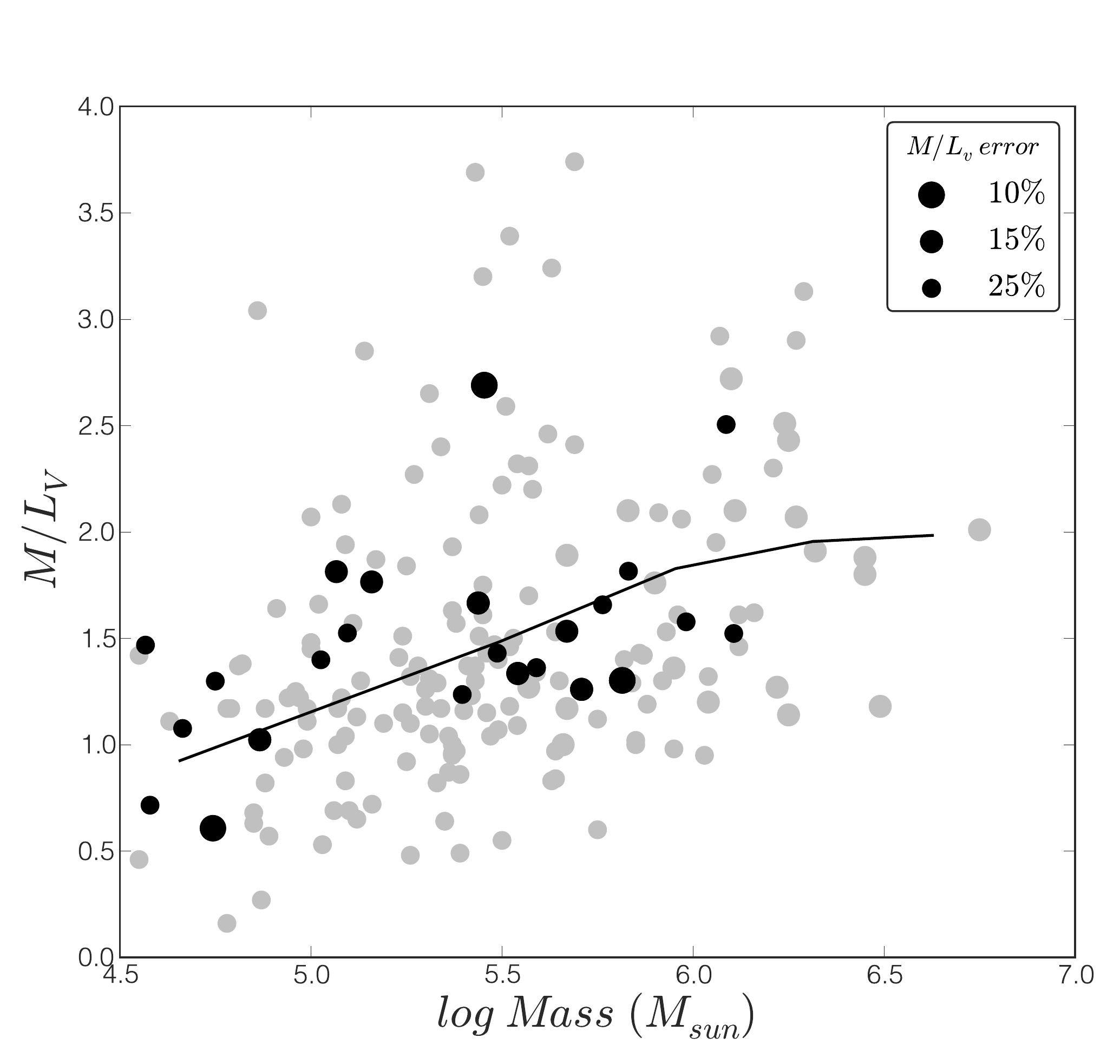}
\caption{$M/L_{v}$ vs. Log Mass for Milky Way GCs in the context of M~31. Gray points correspond to the GCs in M 31 \citep{strader2011}. The point size is inversely proportional to the $M/L_{v}$ error, as indicated in the legend. The black line is the cluster dynamical evolution curve from discussed in \cite{kruijssen2009a}, \cite{kruijssen2008} and \cite{kruijssen2009b} at an Fe/H of -1.3, an age of 12.4 Gyr and a dissolution time of 1 Myr. }\label{fig:masstolightMass}
\end{figure}

In Figure \ref{fig:masstolightMass} we show our data (black points) plotted over the \cite{strader2011} data (gray points) in the mass-$M/L$ plane. Our data show the same trend of increasing $M/L$ with mass as seen in previous work \citep{mandushev1991,strader2011}.  We have also included a dynamical evolution model curve at an [Fe/H]$=-1.3$, an age of 12.4 Gyr and a dissolution time of 1 Myr \citep{kruijssen2008}; the same model was found to qualitatively agree with the M31 GC data \citep{strader2011}.

\subsection{$M/L$ vs.~[Fe/H]}
Due to line blanketing, SSP models with constant IMFs predict higher $M/L$s with increasing metallicity\citep[e.g.][]{bruzual2003,maraston2005}. However, with their large sample of GCs in M~31 \cite{strader2011} found the opposite trend, with lower $M/L$ values at higher metallicity in both V- and K-band.  \cite{strader2011} suggests that the trend may result from a systematic change in the IMF with metallicity.  Specifically, metal-rich clusters would need to have fewer low-mass stars than metal-poor clusters. The best way to see this would be direct counting of low mass stars in MW GCs. Spectroscopic measurements to measure dwarf-to-giant ratios may also be a possible way to verify this result, however, the predicted differences between normal IMFs and the bottom-light IMFs needed to account for the $M/L$s seen in GCs are very small \citep{vandokkum2011}.
 
\begin{figure*}
\plottwo{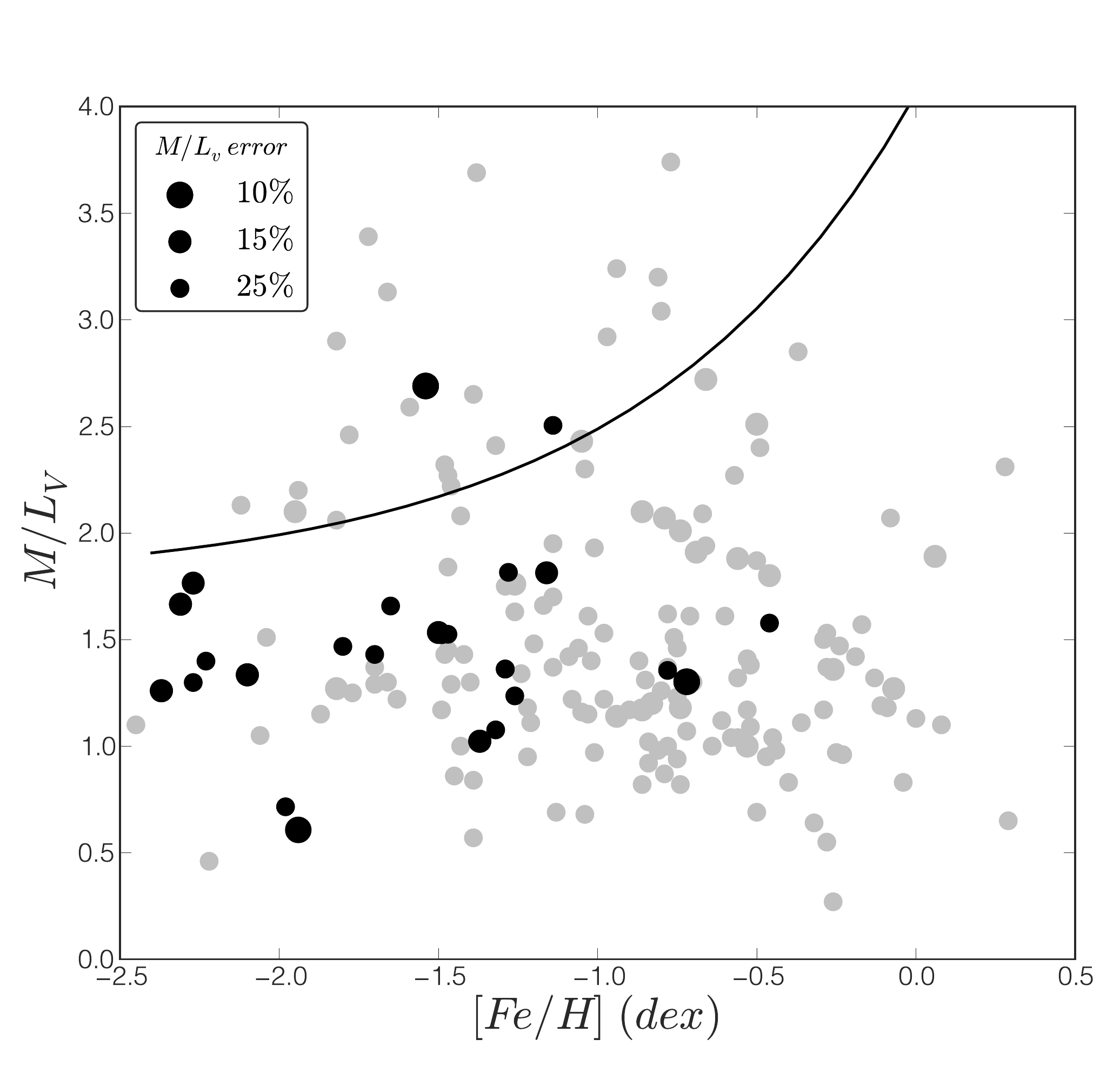}{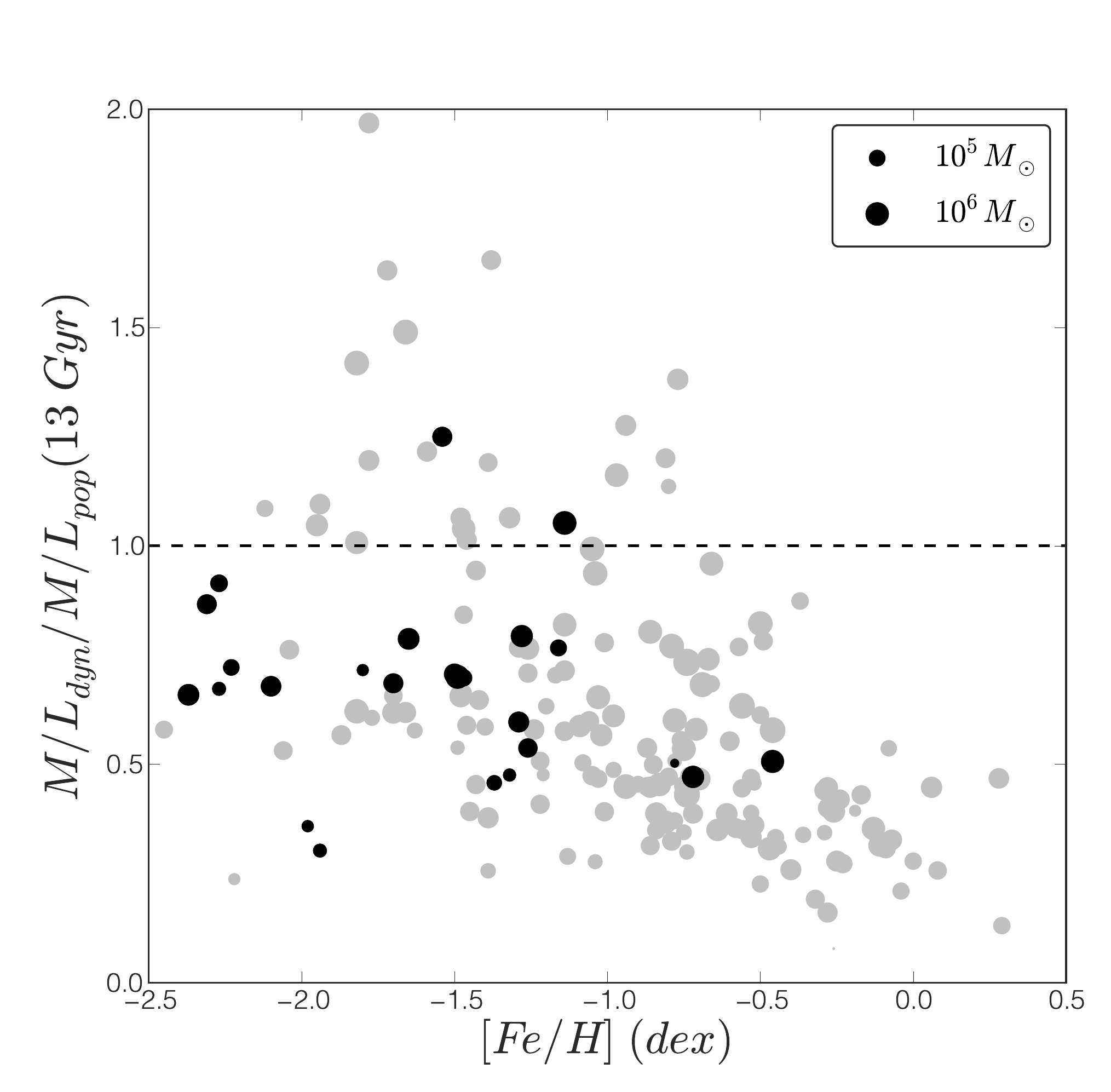}
\caption{{\em Left --} $M/L_{v}$ vs.~metallicity for Milky Way GCs (black points). Gray points correspond to the GCs in M 31 \citep{strader2011}. The point size is inversely proportional to the $M/L_{v}$ error, as indicated in the legend. A stellar population model curve from \cite{mieske2013} for an age of 13 Gyr is overplotted. {\em Right --} The ratio of the derived dynamical $M/L_V$ ($M/L_{dyn}$) to the population model $M/L_{pop}$ curve from \cite{mieske2013} vs.~metallicity.  The points in this panel are proportional to cluster mass, with the sizes given in the legend.  This plot clearly shows that the high metallicity clusters have significantly lower masses than predicted by stellar population models.}
\label{fig:masstolightFeh}
\end{figure*}

In Figure \ref{fig:masstolightFeh} we show the $M/L$ vs. metallicity for the clusters in our sample in black points. We show the M~31 GC values derived by \cite{strader2011} in light gray points. The left panel of Fig.~\ref{fig:masstolightFeh} also shows an $M/L_{pop}$ as a function of [Fe/H] curve from SSP models taken from \cite{mieske2013} for an SSP model with age of 13 Gyr \citep{bruzual2003,maraston2005}.  Using this curve, we divide each clusters' dynamical $M/L$ with the expected $M/L_{pop}$ and show this in the right panel of Fig.~\ref{fig:masstolightFeh}.  Our results are consistent with those of \cite{strader2011}: we see no evidence for the predicted increase in $M/L$ with increasing metallicity.  In fact, at the metal-rich end, both Milky Way and M31 GCs have $M/L$ values $\gtrsim$2$\times$ lower than the expected values from SSP models.   

\subsection{Rotation vs.~[Fe/H]}

\begin{figure}
\centering
\epsscale{1.25}
\plotone{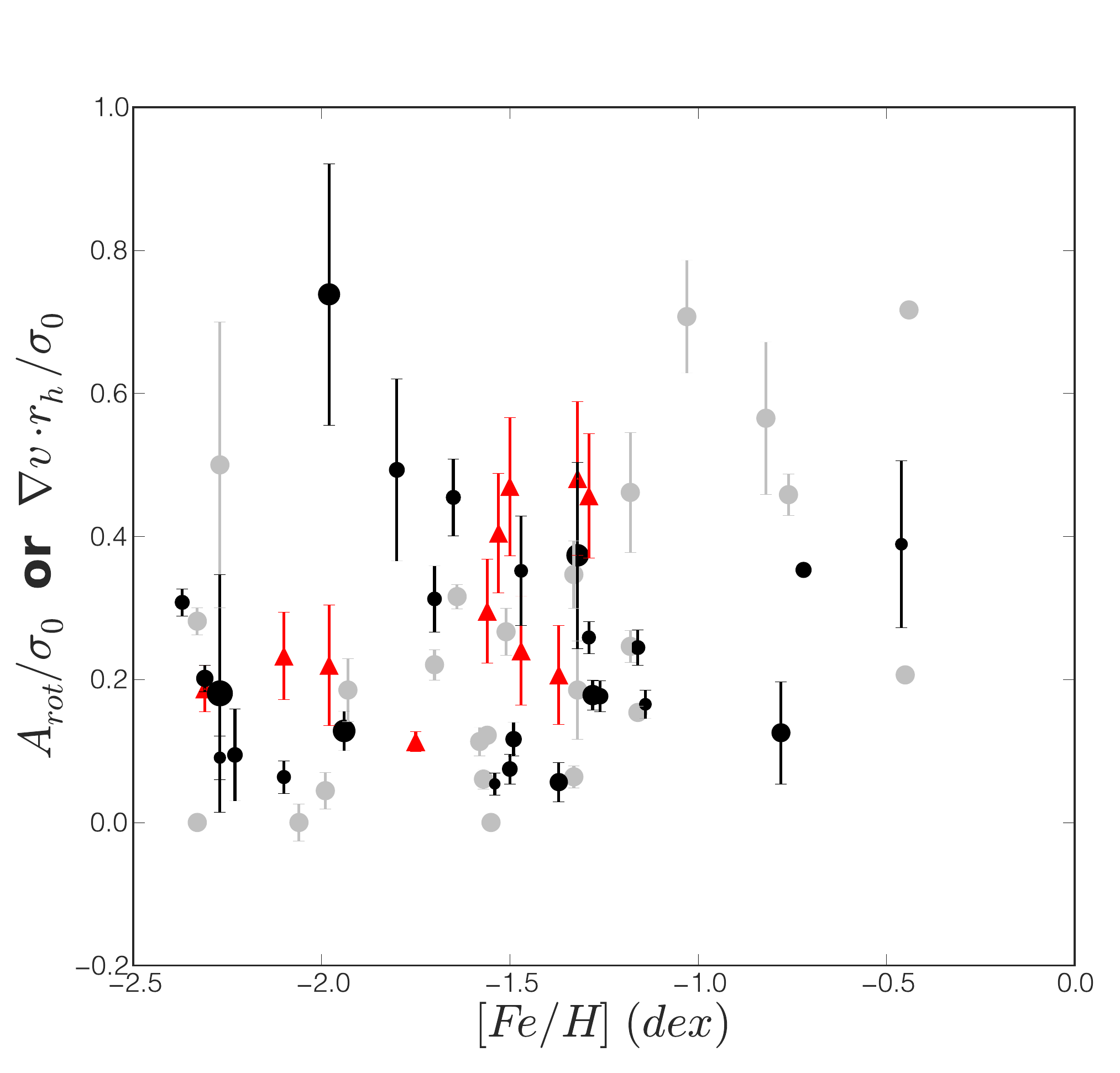}
\caption{Rotations of GCs vs. [Fe/H]. We show the clusters from Table \ref{tab:DispMass} in the Hectochelle, archival data sections, highlighted in the combined data section with black circle. Light gray circles represent the values published in \cite{bellazzini2012}. Red triangles represent data from \cite{fabricius2014}. 
}\label{fig:rotationfe}
\end{figure}

Previous work has shown a trend of increasing rotation with increasing metallicity and more extented horizontal branches \citep{bellazzini2012}. An important consideration when looking for these trends is the radial distribution of the data. Rotation in GCs typically peaks at 1-2 $r_h$ \citep{fiestas2006,bianchini2013}, therefore a GC with high concentrations of data at large radii is less likely to capture the rotational signature that would be seen if the GC was centrally targeted. To characterize the radial coverage of each dataset we included the average of the radius over the effective radius in column 5 of Tab.~\ref{tab:rotation} ($<r/r_{h}>$). In Figure \ref{fig:rotationfe} we show the maximum rotation amplitude divided by the central dispersion ($A_{rot}/\sigma_{0}$) versus metallicity for the GCs in our sample. Due to the radial effect on rotation mentioned above we have scaled the size of points in Fig.~\ref{fig:rotationfe} by $<r/r_{h}>^{-1}$.

Our rotation strengths do not correlate strongly with metallicity like those in \cite{bellazzini2012}. However, the large average radius of stars in both our data and that of \cite{bellazzini2012} suggest that we are not accurately assessing the strength of rotational support in each cluster. For clusters in our sample with centrally concentrated data we derive significatnt rotations, e.g.~in NGC~5466 we find an $A_{rot}/\sigma \sim 0.75$. The velocity gradient seen at the center of cluster may provide a more accurate quantification of the strength of rotation in a cluster \citep{fabricius2014}. The velocity gradient can be used as a proxy for the maximum rotation amplitude ($A_{rot}$) when multiplied by the effective radius \citep{fabricius2014}. Therefore, we include the values from \cite{fabricius2014} in Figure \ref{fig:rotationfe}, by multiplying each velocity gradient by effective radius from \cite{harris1996} (2010 edition) and divided by the central dispersion (using our values, where available, or otherwise using data from \cite{harris1996} (2010 edition)). We see in Fig. \ref{fig:rotationfe} that the red points \citep{fabricius2014} show a larger spread in the rotation strengths for metal-rich GCs than for metal-poor GCs. Overall, we do not see strong evidence for a correlation of metallicity and rotation as seen by \cite{bellazzini2012}.

\section{Conclusions}
\label{sec:conclusions}

In this paper we use kinematic data to derive the dispersions, rotations, masses and mass-to-light ratios for a sample of 25 galactic GCs in a consisten manner.  Below we summarize our primary results.

\begin{itemize}

\item We present velocity measurements with accurate errors for 1951 stars in 12 northern Milky Way GCs from new data taken with the multi-object spectrograph Hectochelle on the MMT \citep{szentgyorgyi2011}. These data are made available in Table \ref{tab:hect_vel}.

\item We fit single-mass King model dispersion profiles to individual stellar velocities in each cluster using a maximum likelihood method that includes a membership determination based on velocity and radial distance.  The binned dispersion profiles of the clusters are remarkably consistent with the King model predictions except for in core-collapsed clusters.  

\item From the best-fit central dispersions we derive $M/L$s and masses for each cluster; this is the largest consistent derivation of these quantities derived solely from resolved stellar data.  Overall, we find good agreement between our measurements and previous measurements.  

\item We confirm the previously observed trend of increasing $M/L$ with cluster mass \citep{mandushev1991,strader2011}. This likely results from dynamical evolution \citep{kruijssen2008}.  

\item We find that $M/L$s do not increase with metallicity as expected from SSP models. This agrees with results previously found in M~31 GCs by \citep{strader2011}.  More specifically, the $M/L$s of metal-rich clusters are $\sim$2$\times$ lower than expected from stellar population models. 

\item We do not find any significant trend of increasing rotation with increasing metallicity as had been previously found by \cite{bellazzini2012}.

\end{itemize}

{\em Acknowledgements:} Observations reported here were obtained at the MMT Observatory, a joint facility of the Smithsonian Institution and the University of Arizona.  We thank Bill Harris for help with his catalog. We'd also like to thank the referee for their comments which improved this paper.

\clearpage
\appendix

\section{Hectochelle Velocities}
\label{sec:vel_appendix}
Here we present all of the velocities derived and used in this study from the multi-object spectrograph Hectochelle.

\begin{deluxetable*}{lcccc}[h]
\tablecolumns{5}
\tablewidth{0pt} 
\tabletypesize{\scriptsize}
\tablecaption{Hectochelle Velocities \label{tab:hect_vel} }  
\tablehead{
\colhead{Cluster} & \colhead{RA} & \colhead{DEC} & \colhead{Velocity} & \colhead{Velocity Error}  \\ 
}
\startdata
M 14 & 17 38 12.84 & -3 06 47.6 & -139.74 & 0.322 \\ 
M 14 & 17 37 02.33 & -3 21 21.4 & 13.942 & 0.116 \\ 
M 14 & 17 37 31.06 & -3 09 26.6 & -52.837 & 0.297 \\ 
.    &   .   &   .   &  .     &    .  \\
\enddata
\end{deluxetable*}

\clearpage
\bibliographystyle{apj}
\bibliography{\myref}
\end{document}